\documentclass[journal]{IEEEtran}
\usepackage{amssymb}
\usepackage[algoruled,vlined,linesnumbered]{algorithm2e}
\usepackage{comment}
\usepackage{amsthm}

\theoremstyle{definition}

\theoremstyle{remark}

\usepackage{bm}
\usepackage{cite} 
\usepackage{amsmath}
\usepackage{float}
\usepackage{graphicx}
\usepackage{subfigure}
\usepackage{xcolor}
\usepackage{booktabs}

\usepackage{amssymb}
\usepackage{multirow}
\usepackage{array}
\newcolumntype{C}[1]{>{\centering\let\newline\\\arraybackslash\hspace{0pt}}m{#1}}

\usepackage{colortbl}
\newcolumntype{A}{>{\columncolor{green!10}}}

\usepackage{diagbox}

\begin{document}
\title{WINNet: Wavelet-inspired Invertible Network for Image Denoising}

\author{Jun-Jie~Huang,
        and~Pier Luigi~Dragotti
        }

\maketitle

\begin{abstract}
Image denoising aims to restore a clean image from an observed noisy image. The model-based image denoising approaches can achieve good generalization ability over different noise levels and are with high interpretability. Learning-based approaches are able to achieve better results, but usually with weaker generalization ability and interpretability. In this paper, we propose a wavelet-inspired invertible network (WINNet) to combine the merits of the wavelet-based approaches and learning-based approaches.
The proposed WINNet consists of $K$-scale of lifting inspired invertible neural networks (LINNs) and sparsity-driven denoising networks together with a noise estimation network. 
The network architecture of LINNs is inspired by the lifting scheme in wavelets. LINNs are used to learn a non-linear redundant transform with perfect reconstruction property to facilitate noise removal. 
The denoising network implements a sparse coding process for denoising. The noise estimation network estimates the noise level from the input image which will be used to adaptively adjust the soft-thresholds in LINNs.
The forward transform of LINNs produce a redundant multi-scale representation for denoising. The denoised image is reconstructed using the inverse transform of LINNs with the denoised detail channels and the original coarse channel. 
The simulation results show that the proposed WINNet method is highly interpretable and has strong generalization ability to unseen noise levels. It also achieves competitive results in the non-blind/blind image denoising and in image deblurring.
\end{abstract}


\IEEEpeerreviewmaketitle

\section{Introduction}

Image denoising is a classical and fundamental inverse problem in image processing and computer vision. Image denoising algorithms aim to restore a noiseless image from noisy observations obtained by digital cameras. 
Given that the observations are inevitably noisy due to the random nature of the photon emission and sensing process, and the imperfection of the signal conversion process \cite{brooks2019unprocessing, wang2020practical}; image denoising is an essential step for further image processing and computer vision applications. With the plug-and-play and the unfolding technique \cite{Venkatakrishnan_P3, Metzler_BM3DAMP, romano2017little, zhang2017learning, meinhardt2017learning}, image denoising algorithms have become even more important as they can also be used to solve other image restoration problems by acting as a powerful prior. 
In this paper, we assume the noise is additive, white and Gaussian. The observed noisy image $\bm{y}$ is expressed as:
\begin{equation}
    \bm{y} = \bm{x} + \bm{n},
    \label{eq:noise}
\end{equation}
where $\bm{x}$ is the clean image and $\bm{n} \sim  \mathcal{N}(0,\sigma^2)$ represents the measurement noise with variance $\sigma^2$.

Image denoising has a rich literature. Depending on how priors have been exploited, image denoising algorithms can be generally classified into model-based methods \textit{e.g.,} \cite{donoho1994ideal, donoho1995adapting, chang2000adaptive, blu2007sure, elad2006image, dong2011sparsity, buades2005non, mahmoudi2005fast, dabov2007image, gu2014weighted, MSWNNW} and learning-based methods \textit{e.g.,} \cite{burger2012image, chen2016trainable, zhang2017beyond, zhang2018ffdnet, Guo2019Cbdnet}. Currently, the learning-based methods achieve state-of-the-art performances, but the model-based methods are still highly competitive.

The model-based methods \cite{donoho1994ideal, donoho1995adapting, chang2000adaptive, blu2007sure, buades2005non, mahmoudi2005fast, elad2006image, dong2011sparsity, dabov2007image, gu2014weighted, MSWNNW} use optimization strategies based on well-defined image priors or noise statistics which lead to algorithms with good interpretability and strong generalization ability. The typical priors used for image denoising include for instance image-domain smoothness, transform-domain sparsity \cite{donoho1994ideal, donoho1995adapting, chang2000adaptive, blu2007sure, elad2006image, dong2011sparsity}, patch-domain non-local self-similarity \cite{dabov2007image, mahmoudi2005fast} and low-rank \cite{gu2014weighted, MSWNNW}. 


The wavelet transform has been very effective in many imaging applications due to its ability to provide sparse representation of images.
It provides a versatile multi-resolution analysis with perfect reconstruction property and time-frequency localization property. Due to its properties, the wavelet transform has been applied for a wide range of image restoration problems, including image denoising \cite{donoho1994ideal, chang2000adaptive, blu2007sure}, image deconvolution \cite{donoho2004fast, pustelnik1999wavelet} and image inpainting \cite{dong2012wavelet, he2014iterative}.
The wavelet transform is a fixed transform and, in some cases, learning a transformation better adapted to the data at hand may lead to more effective solutions. Dictionary learning tries to achieve that by learning a (redundant) sparsifying transform from training data.
Elad and Aharon \cite{elad2006image} proposed to perform image denoising with learned redundant dictionaries. 
Compared to fixed transforms, the combination of learned dictionaries and sparse coding leads to significantly improved denoising performance. 
However, both the analytical transforms and the learned dictionaries are linear transformations. A suitable non-linear transform with perfect reconstruction property has the potential to achieve better performances.

Besides the nice properties of the wavelet transform, the noise adaptive non-linear operator also contributes to the effectiveness of wavelet-based methods.
Donoho and Johnstone \cite{donoho1994ideal} proposed soft-thresholding operator and applied it with the optimal ``universal threshold" $T=\sqrt{2 \sigma^2 \log N}$ (where 
$N$ is the number of samples) to the wavelet-domain coefficients to remove noise. Chang \textit{et al.} \cite{chang2000adaptive} proposed a BayesShrink threshold $T = \hat{\sigma}^2 / \hat{\sigma}_X$ (where $\hat{\sigma}$ and $\hat{\sigma}_X$ are the estimated standard deviation of noise and signal, respectively) for soft-thresholding with a Bayesian framework for wavelet-domain image denoising. 
Both the ``universal threshold" and BayesShrink threshold are adaptively adjusted for different noise levels.

The learning-based methods \cite{burger2012image, chen2016trainable, zhang2017beyond, zhang2018ffdnet, Guo2019Cbdnet} construct a denoising model by learning from noisy-clean image pairs. 
The availability of training data and the flexibility of network structure simplifies and enriches the algorithm design and boosts the performance, while the learned models have more restricted generalization ability compared to the model-based methods and are usually treated as black-box systems with poor interpretability. 
Therefore, learning-based methods mainly aim to explore more efficient and effective architecture as well as strategies to improve the generalization ability of the learned denoising model.
The denoising convolutional neural networks (DnCNN) method \cite{zhang2017beyond} learns a deep CNN model with batch normalization layers and a skip connection. By learning from training samples with different noise levels, DnCNN can perform blind image denoising when the noise level of the testing image is within the range of the training noise levels. The fast and flexible denoising network (FFDNet) \cite{zhang2018ffdnet} takes the noise level map and the noisy image as input to the model, and therefore can handle spatially varying noises. 
The convolutional blind denoising network (CBDNet) \cite{Guo2019Cbdnet} consists of two convolution subnetworks: a noise estimation subnetwork and a non-blind denoising subnetwork. The noise estimation subnetwork learns to infer a noise level map from the noisy image, and the non-blind denoising subnetwork which has a U-Net like structure takes both the noise level map and input noisy image as input and estimates the clean image. 

Here we propose to learn a (redundant) invertible sparsifying transform by leveraging the invertible neural network framework. In our design we want our non-linear transform to inherit some of the properties of the wavelet transform, including its sparsifying ability as well as its multi-scale property where wavelets at larger scales are more global that the wavelets at finer scales which are instead very local.
We aim to combine the merits of the model-based and learning-based image denoising approaches. Specifically by following a strategy similar to wavelet domain thresholding, we aim to enhance the generalization ability and interpretability of the learning-based image denoising method.

We propose a novel wavelet-inspired invertible network (WINNet) for image denoising. The architecture is shown in Fig. \ref{fig:overview} and is described in detail in Section \ref{sec:overview} and \ref{sec:LINN}.
Rather than learning features with unconstrained CNNs,
we propose to learn a non-linear wavelet-like transform with perfect reconstruction property using invertible neural networks with a structure inspired by the lifting scheme \cite{sweldens1998lifting, daubechies1998factoring}. We call these lifting inspired networks LINNs. With PR property, the learned LINNs can serve as a versatile transform.
For denoising, a sparsity-driven denoising network is applied to the transform coefficients to remove the noise. The restored image can then be reconstructed using the inverse transform of the learned LINNs.

The proposed WINNet is made of several LINNs, one per scale. Moreover, to achieve good generalization ability, all the soft-thresholds in WINNet are set to be noise adaptive and can be adjusted according to the estimated noise level. In this way, the proposed denoising network achieves good generalization ability even to unseen noise. The noise level is estimated using a model-inspired noise estimation network which exploits low-rank patches on the input noisy image and estimates noise levels as the minimum singular value of the weighted patches.

The contribution of this paper is three-fold:
\begin{itemize}
    \item We propose an invertible thresholding network for image denoising. It is designed based on the principles of wavelet-based methods, therefore leads to a model with high interpretability. The LINNs at various scales produce a non-linear redundant transform with perfect reconstruction property, and the denoising network implements a basis pursuit denoising process. 
    
    \item The proposed method is able to achieve blind image denoising. The noise estimation network estimates noise from the input image. To adapt the learned WINNet to the test noise level, the soft-thresholds in network are adaptively adjusted with respect to the estimated noise level.
    
    \item The proposed WINNet is with high interpretability and strong generalization ability. It achieves performances comparable to those of state-of-the-art algorithms on non-blind/blind image denoising and image deblurring.
\end{itemize}


The rest of the paper is organized as follows: Section \ref{sec:relatedworks} discusses related works in image denoising and invertible neural networks. Section \ref{sec:overview} provides an overview of WINNet. Section \ref{sec:LINN} describes the architecture of LINNs which are the building blocks of WINNet, whereas Section \ref{sec:UBD} focuses on the denoising network, the noise estimation strategy and the training strategy. Section \ref{sec:results} shows the numerical results, compares the proposed method with other image denoising methods and demonstrates the application of WINNet on image deblurring. Finally Section \ref{sec:conclude} concludes the paper.

\section{Related Work} \label{sec:relatedworks}

\subsection{Learning Universal Image Denoising}

Blind image denoising can be achieved by learning from training samples with varying noise levels \cite{zhang2017beyond, zhang2018ffdnet, Guo2019Cbdnet}, however, the learned deep models still have poor generalization ability on images with noise levels beyond the training levels. Hence, one may wish to learn a universal denoising model with good performances on arbitrary noise levels.

Mohan \textit{et al.} \cite{mohan2020BFCNN} found that the bias term in a denoising CNN will lead to a model bias to the training noise levels. Hence, they propose to remove all the bias terms in both ReLU activation functions and the batch normalization layers. The resultant bias-free CNN (BF-CNN) has a scale homogeneity property, \textit{i.e.}, $f_{\text{BF}}(\alpha \bm{y}) = \alpha f_{\text{BF}}(\bm{y})$ where $f_{\text{BF}}(\cdot)$ denotes a bias-free CNN and $\alpha>0$, and shows consistently better generalization ability than its counterpart with bias terms. 
Helou and Süsstrunk \cite{Helou2020BUIFD} proposed a blind universal image fusion denoiser (BUIFD) whose structure is derived from a Bayesian framework and consists of a noise level CNN, a prior CNN and a fusion network. The information of noisy image and image prior are fused based on signal-to-noise ratio (SNR). By explicitly incorporating the noise level, BUIFD shows stronger generalization ability than that of conventional denoising CNNs.

\begin{figure}[t]
    \centering
    
    \hspace*{\fill}
    \subfigure[One level of the analysis filter bank in the wavelet decomposition.]{
		\includegraphics[width=0.45\textwidth]{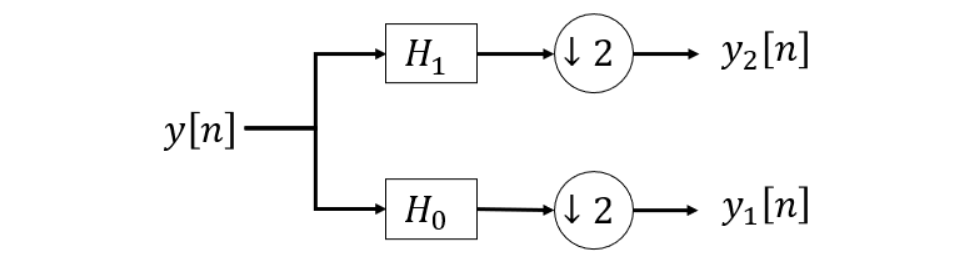}
		\label{fig:WT}
	}
    \hspace*{\fill}
    
    \hspace*{\fill}
    \subfigure[The lifting scheme.]{
		\includegraphics[width=0.45\textwidth]{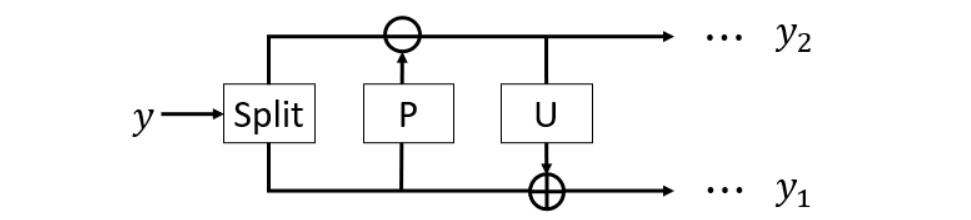}
		\label{fig:LS}
	}
	\hspace*{\fill}
	
	\caption{The wavelet transform obtained using a filter bank and the lifting scheme.}
    \label{fig:WTLS}
\end{figure}

\subsection{Lifting Scheme and Invertible Neural Networks}

The lifting scheme introduced in \cite{sweldens1998lifting, daubechies1998factoring} proposes to construct a wavelet transform by first splitting the signal into an even and an odd part. A predictor is used to predict the odd signal from the even part, therefore leading to a sparse residual signal. The update step is used to adjust the even signal based on the prediction error of the odd part to make it a better coarse version of the original signal. There can be multiple pairs of predictor and updater. The lifting scheme can represent a transform with perfect reconstruction condition and any intermediate representation can be used to infer the input signal and the final representation.
\begin{figure*}
    \centering
    \includegraphics[width=0.80\textwidth]{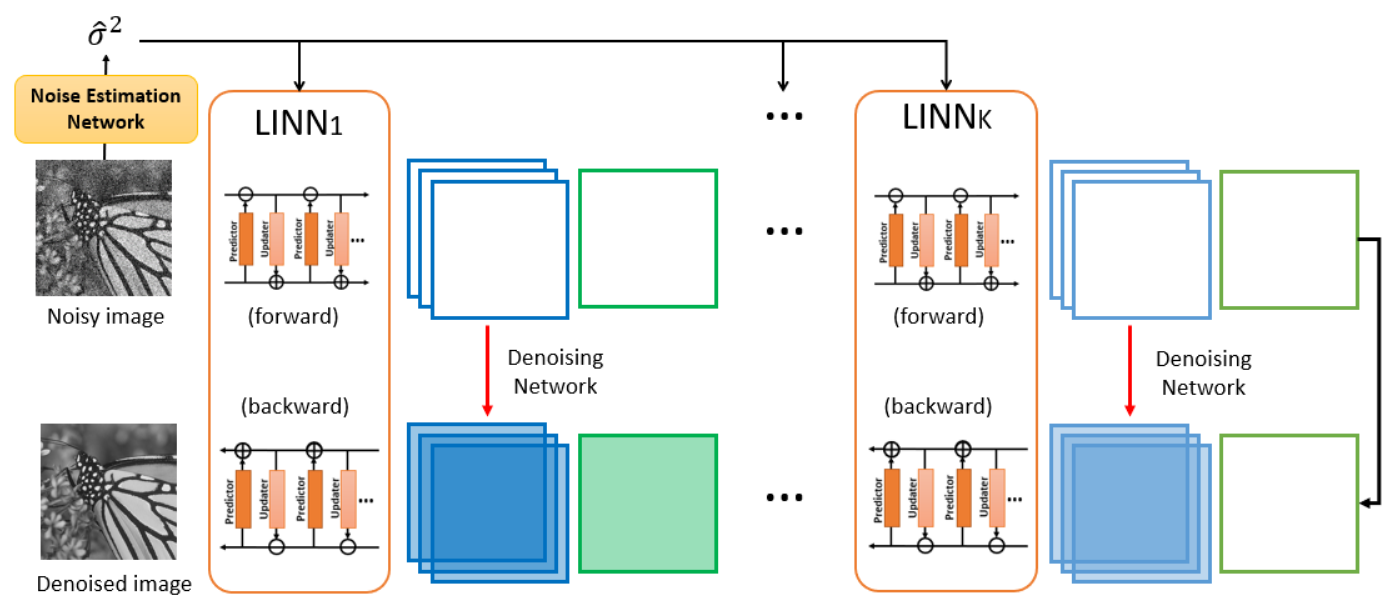}
    \caption{Overview of the proposed wavelet-inspired invertible network (WINNet). It consists of $K$ levels of lifting inspired invertible neural networks (LINN) and denoising network. 
    The forward transform of LINN non-linearly converts the input noisy image into coarse part (green) and detail parts (blue). Denoising network will perform denoising operation on the detail part while the coarse version is decomposed again using a second level and the decomposition and denoising step are repeated $K$ times . The backward transform of the LINN will reconstruct the denoised image using the denoised detail parts and the original coarse part.
    The estimated noise level from the noise estimation network will be used to adjust the soft-thresholds of the soft-thresholding non-linearity to make the WINNet to adapt well to the current noise level.}
    \label{fig:overview}
\end{figure*}

Inspired by this scheme, the invertible neural networks (INNs) \cite{dinh2014nice, dinh2016density, gomez2017reversible, jacobsen2018revnet} was proposed for memory-efficient backpropagation model and used for constructing flow-based generative models. 
INNs are bijective function approximators and the intermediate feature representation of INNs can be used to perfectly reconstruct the input signal.
Different invertible architectures have been proposed, including coupling layer \cite{dinh2014nice}, affine coupling layer \cite{dinh2016density},  reversible residual network \cite{gomez2017reversible} and i-RevNet architecture \cite{jacobsen2018revnet}.
INNs have been applied for low-level computer vision tasks, including image rescaling \cite{xiao2020invertible}, image segmentation \cite{Etmann2020iUnet}, image colorization \cite{ardizzone2019guided, Rui21Decolor}, image compression \cite{Ma2020Compression, Ma2020iWave, Li2020REVAE}, and image denoising \cite{Huang2020LINN, liu2021invertible}. 

\section{Overview} \label{sec:overview}

Our goal is to design an invertible neural network (INN) where the design choices are made to induce an invertible transform with properties similar to the wavelet transform.
In Fig. \ref{fig:WT}, we show one level of a 1-D discrete wavelet transform which is obtained using a filter bank. Here $y_1[n]$ is a ``smooth" version of $y$ and $y_2[n]$ contains high-frequency information of $y[n]$ and is sparse. We know that this decomposition can be achieved using the lifting scheme shown in Fig. \ref{fig:LS}.

The split operation divides the sequence into the even and odd components, ``P" predicts the even using the odd leading to a sparse error sequence and ``U" updates the odd sequence to make it smoother. The process is then repeated until $y_1$ and $y_2$ are obtained. ``P" and ``U" are fixed and depend on the filters $H_0$ and $H_1$. A two-level wavelet decomposition can also be implemented with the lifting scheme by applying it again on $y_1[n]$.
We note that the equivalent low-pass filter in the two-scale case is $H_0(z)H_0(z^2)$ and is less local than $H_0(z)$. It is therefore more adapted at catching more global proprieties of $y$. The transforms discussed so far are non-redundant but the undecimated wavelet transform can be easily obtained by shifting-out the downsampler in Fig. \ref{fig:WT} and by upsampling the filter coefficients by $2^{j-1}$, with $j=1, \cdots, J,$ and $J$ is the number of decomposition levels. The upsampling is essential to keep the multi-resolution property of the wavelet transform. 

The architecture of our INN is inspired by the lifting scheme in the wavelet transform and the multi-resolution property of the wavelet transform. A WINNet with $K$-level lifting inspired neural network (LINN) is shown in Fig. \ref{fig:overview}, where we iterate the decomposition only on the coarse version of the image and replace the standard ``P" and ``U" operations with two convolutional neural networks. Through the use of a denoising strategy we enforce ``P" to act as a ``Predictor" and ``U" as an ``Update". Moreover, at larger scales, we upsample the filters in the convolutional layers to mimic the algorithm \`{a} trous \cite{shensa1992discrete} used in wavelets theory to implement the wavelet transform. Finally we replace the split operator with a redundant invertible linear operator which leads to a redundant decomposition. 

We achieve sparsity by applying a non-invertible denoising operation on the detail coefficients. The denoising operation forces the network to sparsify the detail coefficients so that only noise is removed. The network we use for denoising is also a sparsity-driven network. WINNet is then trained using noisy images as input and ground-truth noiseless images as output. By training the network with paired noisy and clean images, we force the network to sparsify the detail coefficients whilst preserving essential information in the coarse part. Besides leading to an effective and versatile non-linear invertible transform, this approach also leads to a very effective denoising algorithm. To make it more universal we also include a network to estimate the noise variance which in turn changes the soft-thresholds in WINNet. In this way, our system adapts automatically to different unknown levels of noise. We also note that our scheme is highly interpretable as we know what each part of the network is doing. 
In what follows we describe in more detail each element of our network.

\section{Lifting Inspired Invertible Neural Networks} 
\label{sec:LINN}

\begin{figure}[t]
    \centering
    
    \subfigure[The forward transform of a LINN with splitting operator $\mathcal{S} (\bm{y}^{k}, \bm{K}_s )$.]{
		\includegraphics[width=0.36\textwidth]{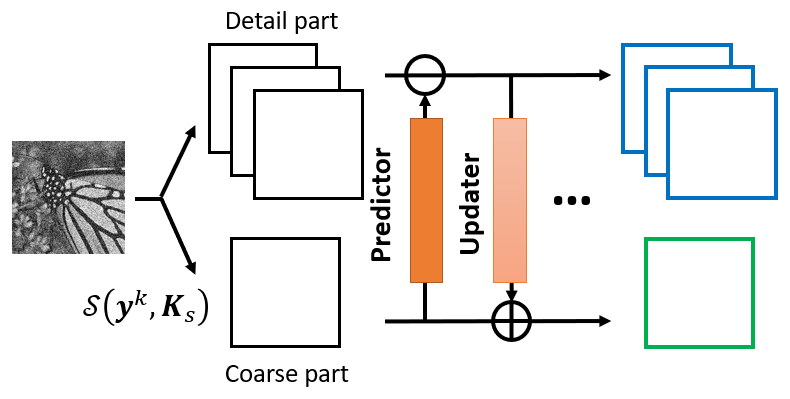}
		\label{fig:LINN_f}
	}
    \hfill
    \subfigure[The inverse transform of a LINN with merging operator $\mathcal{M} (\hat{\bm{f}}^{k}, \bm{K}_s )$.]{
		\includegraphics[width=0.36\textwidth]{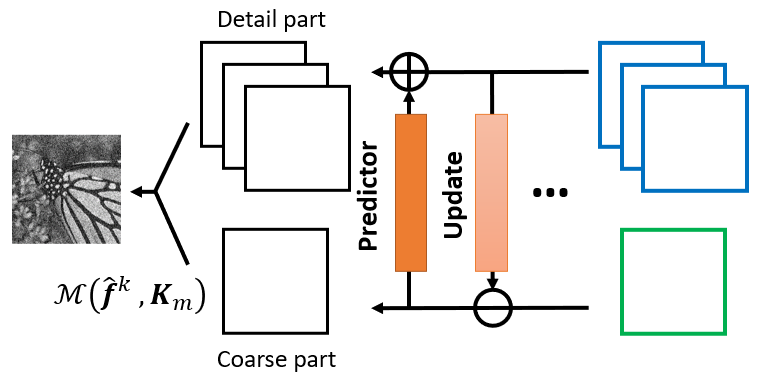}
		\label{fig:LINN_b}
	}

    \caption{The forward transform and the inverse transform of a LINN.}
    \label{fig:LINN}
\end{figure}

The lifting inspired invertible neural network (LINN) is used to replace one level of the wavelet transform (see Fig. \ref{fig:overview} and Fig. \ref{fig:LINN}). Compared to wavelet transforms and learned dictionaries, LINN can represent more complex features.
Similar to the lifting scheme which splits the input signal and then alternates prediction and update \cite{sweldens1998lifting, daubechies1998factoring},
LINN consists of splitting/merging operators and learnable predict and update networks.

Since LINN satisfies PR conditions, the obtained transform coefficients contain the same amount of information as the input image. Though noise is not removed in the transform coefficients, the non-linear transform will learn to separate signal and noise into the coarse and detail parts through the denoising operation applied on the detail part.
The PR property of LINN also makes it a versatile transform for image denoising with different noise levels and for other image restoration problems.

\subsection{Splitting/Merging Operator} 

\label{sec:splitOp}

The splitting operator is used to separate the input image into two parts, and the merging operator performs the inverse of the splitting operator. The splitting/merging operator in current methods \cite{dinh2014nice, dinh2016density, gomez2017reversible, jacobsen2018revnet, Rodriguez2020AdapWavelet, ardizzone2019guided, Ma2020Compression, xiao2020invertible, Etmann2020iUnet} keep the same number of input and output coefficients. 

In this paper, we propose to learn LINNs with redundant representations in analogy with the redundant wavelet transform which provides better performance compared to the decimated wavelet transform in image restoration tasks \cite{Starck2007UndeciWavelet, blu2007sure}. Therefore, the splitting operator will lead to a redundant representation, and merging operator recovers the input image from the redundant representation.

For an input image $\bm{y}^{k} \in \mathbb{R}^{1 \times N_1 \times N_2}$ at the $k$-th scale, the splitting operator $\mathcal{S}$ is parameterized by a convolutional kernel
$\bm{K}_s \in \mathbb{R}^{c \times 1 \times p \times p}$
where $c$ denotes the number of channels and $p$ denotes the spatial filter size. 
A convolution is performed to obtain multi-channel features $\bm{f}^{k} \in \mathbb{R}^{c \times N_1 \times N_2}$:
\begin{equation}
    \bm{f}^{k} = \bm{K}_s \otimes \bm{y}^{k},
\end{equation}
where $\otimes$ denotes the convolution operator.

The split operation then divides the first $h$ channels of $\bm{f}^{k}$ into coarse part $\bm{c}^{k} \in \mathbb{R}^{h \times N_1 \times N_2}$, and the remaining $c-h$ channels into detail part $\bm{d}^{k} \in \mathbb{R}^{(c - h) \times N_1 \times N_2}$. In this paper, we set the number of coarse channels to $h=1$,
and we denote the splitting operator as:
\begin{equation}
    \mathcal{S} (\bm{y}^{k}, \bm{K}_s ) = ( \bm{c}^{k}, \bm{d}^{k} ).
\end{equation}

The merging operator $\mathcal{M}$ represents the inverse of the splitting operator, and is parameterized by a convolutional kernel 
$\bm{K}_m \in \mathbb{R}^{1 \times c \times p \times p}$. 
It first concatenates $\bm{c}^{k}$ and $\bm{d}^{k}$ and this results in $\hat{\bm{f}}^{k}$. A convolution is performed to recover $\bm{y}^{k}$:
\begin{equation}
    \bm{y}^{k} = \bm{K}_m \otimes \hat{\bm{f}}^{k}.
\end{equation}

The merging operator can then be denoted as:
\begin{equation}
    \mathcal{M} (\hat{\bm{f}}^{k}, \bm{K}_m ) = \bm{y}^{k}.
\end{equation}

To achieve invertibility, we can pick $\bm{K}_s$ so that it can be reshaped to a tight frame with $c \geq p^2$. The merging convolutional kernel then simply corresponds to the transpose of the splitting convolutional kernel. This will ensure the splitting and merging operators are invertible. We can use orthogonal transforms of size $c \times c$ such as orthogonal wavelet transforms, Discrete Cosine Transforms (DCT) and other analytical transforms as the splitting and merging operators. The tight frame is then obtained by picking $p^2$ columns of the original orthogonal matrix.

Besides the analytical transforms, it is also possible to learn an orthogonal matrix with proper parameterization methods, for example, the Cayley transform \cite{absil2009optimization} of a skew-symmetric matrix $\bm{\Theta} - \bm{\Theta}^T$ is an orthogonal matrix:
\begin{equation}
    \bm{K} = \left(\mathrm{I} - (\bm{\Theta} - \bm{\Theta}^T)\right) \left(\mathrm{I} + (\bm{\Theta} - \bm{\Theta}^T)\right)^{-1},
\end{equation}
where $\mathrm{I}$ is the identity matrix. The corresponding tight frame $\bm{K}_s$ can therefore be a sub-matrix (\textit{i.e.}, extracting $p^2$ columns) of the learned orthogonal matrix $\bm{K}$ parameterized by learnable parameters $\bm{\Theta} \in \mathbb{R}^{c \times c}$.



\begin{figure}
    \centering
    \subfigure[A depth-wise separable convolution layer is composed of a depth-wise convolution layer and a $1 \times 1$ convolution layer.]{
        \includegraphics[width=0.4\textwidth]{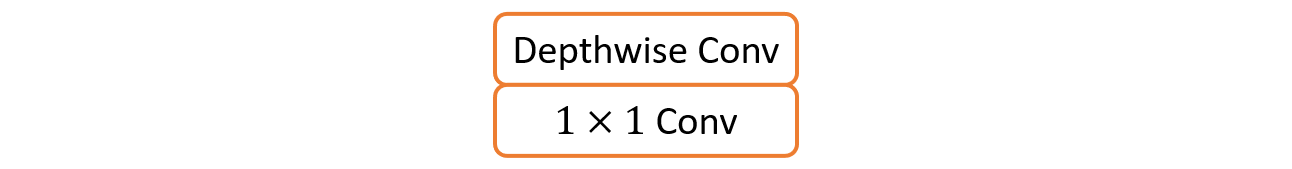}
        \label{fig:SepConv}
    }
	\hfill
	\subfigure[The residual block with separable convolution layer and soft-thresholding non-linearity.]{
		\includegraphics[width=0.4\textwidth]{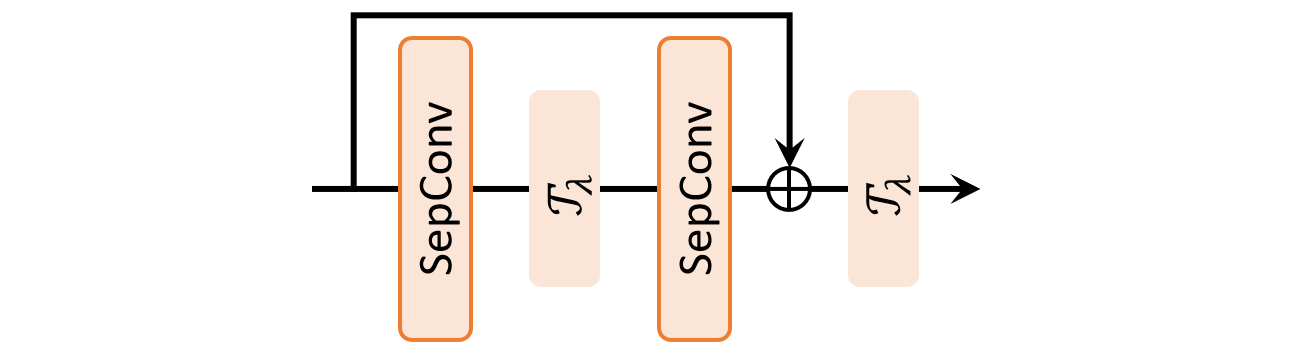}
		\label{fig:ResBlock}
	}
	\hfill
	\subfigure[The network architecture of the Predict/Update network.]{
		\includegraphics[width=0.4\textwidth]{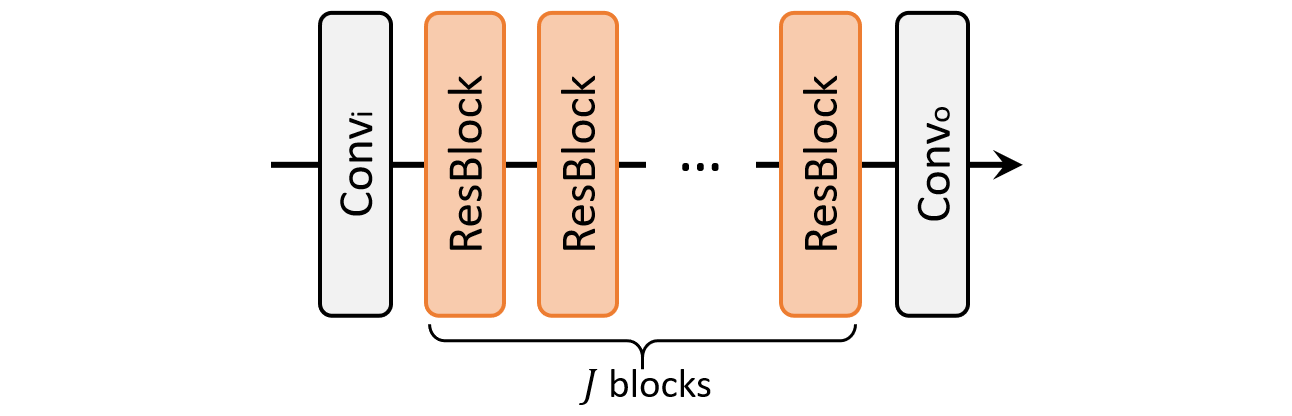}
		\label{fig:PUNet}
	}
    \caption{The building blocks and network structure of the PUNets}
    \label{fig:ResNet_SepConv}
\end{figure}
\subsection{Network Architecture of LINN} 


At each scale, the Predict/Update networks are shared among the forward and inverse transform of LINN, but are connected with different signs and direction.

Fig. \ref{fig:LINN_f} and Fig. \ref{fig:LINN_b} shows the schematics of the forward transform and the inverse transform of the $k$-th scale LINN, respectively. 
In the forward transform, $\left( \bm{c}^{k}, \bm{d}^{k} \right)$ will be non-linearly transformed to a representation which is more suitable for denoising. After denoising, the denoised detail part and the original coarse part will be transformed back to the original domain using the inverse transform of LINN.

In the forward transform (shown in Fig. \ref{fig:LINN_f}), a Predict network conditioned on the coarse part aims to predict the detail part to make the resultant residual of the detail part sparse. The Update network conditioned on the detail part is used to adjust the coarse part to make it a smoother version of the input image. 
The Predict and Update networks are applied alternatively to update the coarse and detail parts.
Let us denote $\left( \bm{c}^{k}_0, \bm{d}^{k}_0 \right) = \left( \bm{c}^{k}, \bm{d}^{k} \right)$.
The $m$-th pair ($m \in [1,M]$) of update and predict operation can be expressed as:
\begin{align} 
    \bm{d}^{k}_m &=  \bm{d}^{k}_{m-1} - P_m^k \left(\bm{c}^{k}_{m-1} \right), \\ 
    \bm{c}^{k}_m &=  \bm{c}^{k}_{m-1} + U_m^k \left(\bm{d}^{k}_m \right),
    \label{eq:LINNforward}
\end{align}
where $\bm{d}^{k}_m$ and $\bm{c}^{k}_m$ denotes the updated detail part and coarse part using the $m$-th Predict network $P_m^k(\cdot)$ and Update network $U_m^k(\cdot)$, respectively.

In the inverse transform of the $k$-th scale LINN (shown in Fig. \ref{fig:LINN_b}), $\left(\bm{d}^{k}_{m-1}, \bm{c}^{k}_{m-1} \right)$ for $m \in [1, \cdots, M]$ can be estimated based on $\left(\bm{d}^{k}_m, \bm{c}^{k}_m \right)$ and $P_m^k(\cdot)$ and $U_m^k(\cdot)$ as follows:
\begin{align} 
    \bm{c}^{k}_{m-1} &=  \bm{c}^{k}_m - U_m^k \left(\bm{d}^{k}_m \right),\\
    \bm{d}^{k}_{m-1} &=  \bm{d}^{k}_m + P_m^k \left(\bm{c}^{k}_{m-1} \right).
    \label{eq:LINNbackward}
\end{align}

When no lossy operation is applied on $\left( \bm{c}^{k}_M, \bm{d}^{k}_M \right)$, the inverse transform of LINN can, by construction, perfectly recover the inputs of the forward transform with any Predict and Update networks. In this paper, we enforce LINN to learn a sparsifying transformation through denoising.

\subsection{Predict/Update Networks}

The Predict and Update networks can be any functions and their properties will not affect the invertibility of LINN. In this paper, we use the same structure for each Predict/Update networks which we therefore denote with PUNets. 
To achieve a better complexity and performance trade-off, we construct PUNets with depth-wise separable convolution (SepConv) layers \cite{chollet2017xception} and soft-thresholding non-linearity \cite{donoho1994ideal}. As shown in Fig. \ref{fig:PUNet}, 
a PUNet consists of an input convolutional layer $\text{Conv}_i$ with a soft-thresholding, $J$ residual blocks with SepConv layers, and an output convolutional layer $\text{Conv}_o$. 

\textbf{Separable Convolution:} A depth-wise separable convolution layer \cite{chollet2017xception} consists of a depth-wise convolution layer and a $1 \times 1$ convolution layer (shown in Fig. \ref{fig:SepConv}).
It can reduce the number of parameters, especially when the channel number and the spatial filter size are large.
Compared to the standard convolution of size $c_{out} \times c_{in} \times q \times q$, the SepConv layer only requires $c_{in} \times q \times q + c_{out} \times c_{in}$ parameters.

\textbf{Soft-thresholding:} The soft-thresholding operator is used as the non-linearity in PUNets. The soft-thresholding operator has a tight connection with sparse coding. It can also be considered as a two-sided ReLU non-linearity and can be more effective than ReLU for some image processing applications. The soft-thresholding operator
can be expressed as:
\begin{equation}
    \mathcal{T}_{\bm{\lambda}}(\bm{x}) = \text{sgn}(\bm{x})\max(|\bm{x}|-\bm{\lambda},0).
\end{equation}

To ensure $\bm{\lambda}$ are non-negative, a Softplus function\footnote{$\text{Softplus} (x) = \frac{1}{\beta} \cdot \log (1 + \exp{(\beta \cdot x)})$.} is applied on each learned thresholds. 



\begin{figure}
    \centering
    \hspace*{\fill}
    \subfigure[Dilation level 1.]{
        \includegraphics[width=0.11\textwidth]{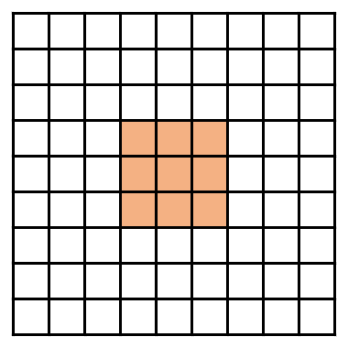}
        \label{fig:dilate1}
    }
	\hspace*{\fill}
	\subfigure[Dilation level 2.]{
		\includegraphics[width=0.11\textwidth]{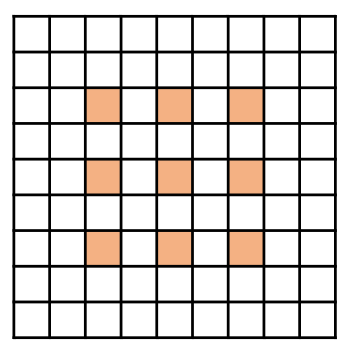}
		\label{fig:dilate2}
	}
    \hspace*{\fill}
    \caption{The dilated filter at different dilation level.}
    \label{fig:atrous}
\end{figure}

\textbf{Multi-scale Property:} Multi-resolution signal decomposition is an essential property of the wavelet transform.
For an input image, the wavelet transform provides a multi-scale analysis which captures the information at different scales.
In order to mimic the wavelet transform, we iterate the LINN-based decomposition on the coarse part. Moreover, we apply \`{a} trous convolution with dilation rate $2^{k-1}$ for the $k$-th scale LINN even though the redundancy factor in our setting is not two. As a result, the larger scale LINN will have a bigger receptive field. Fig. \ref{fig:atrous} shows the exemplars of dilated $3 \times 3$ filter at two different levels.

\section{Blind Image Denoising}
\label{sec:UBD}

The denoising network aims to denoise the detail part $\bm{d}^{k}_M$ by removing the noise features while retaining the image content. The denoising networks are the only non-invertible components in WINNet, and the noisy information can only be removed there. The denoising networks also force LINNs to be a sparsifying transformation which is desirable for many image processing tasks.

\subsection{Sparsity-Driven Denoising Network}
\label{sec:SDNet}

Any existing image denoiser can be applied here, however, as our objective is to achieve an image denoising algorithm with high interpretability, we propose to use a sparsity-driven denoising architecture.

Let us assume that the learned forward transform by LINN is a sparsifying transform and the small number of large coefficients correspond to the image contents. (The forward transformed images by LINN will be shown in Section \ref{sec:results} Simulation Results to validate this assumption).
The denoising process can then be modeled as a basis pursuit denoising problem.
To enrich the representation ability of the sparsity-driven denoising network, we over-parameterize with a convolutional dictionary as:
\begin{equation}
    \bm{g} = \arg \underset{\bm{g}}{\min} \frac{1}{2} \Vert \bm{d}^{k}_M - \sum_{i=1}^N \bm{D}_i \otimes \bm{g}_i \Vert_{2}^{2} + \sum_{i=1}^N \lambda_i \Vert \bm{g}_i \Vert_{1},
    \label{eq:l1-over}
\end{equation}
where $\bm{g} = [\bm{g}_1, \cdots, \bm{g}_N]$ are the sparse features, $\bm{D}=[\bm{D}_1, \cdots, \bm{D}_N]$ is the over-parameterized convolutional dictionary, and $\bm{\lambda} = [\lambda_1, \cdots, \lambda_N]$ are the regularization parameters.


	

\begin{figure}
    \centering
    \includegraphics[width=0.48\textwidth]{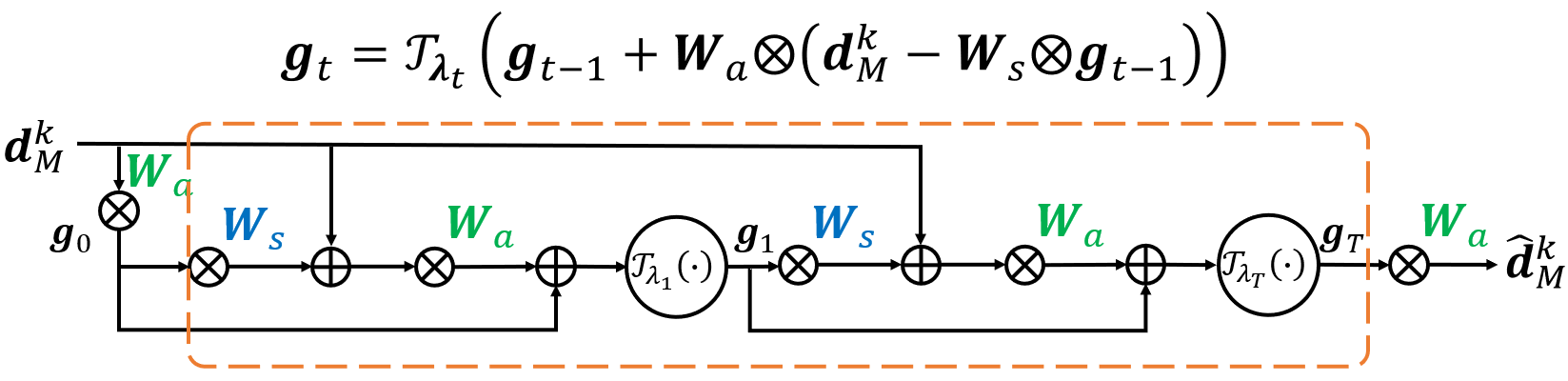}
    \caption{The CLISTA denoising network. At the $t$-th layer, the sparse feature $\bm{g}_{t}$ is estimated using a soft-thresholding operator. The denoised detail channel is estimated from $\bm{g}_T$ with the synthesis convolutional layer.}
    \label{fig:LISTANet}
\end{figure}





The above $l_1$-norm minimization problem can be solved using Iterative Shrinkage-Thresholding Algorithm (ISTA) \cite{daubechies2004iterative}:
\begin{equation}
    \bm{g}_{t} = \mathcal{T}_{\bm{\lambda}/\mu}\left( \bm{g}_{t-1} + \frac{1}{\mu} \bm{D}^T \otimes \left(\bm{d}^{k}_M - \bm{D} \otimes \bm{g}_{t-1} \right) \right),
\end{equation}
where $\mu$ is the step size.



The LISTA \cite{gregor2010learning} algorithm parameterizes the unknown dictionaries in ISTA as learnable parameters.
We apply a convolutional LISTA (CLISTA) as our sparsity-driven denoising network. The schematic of the CLISTA denoising network is shown in Fig. \ref{fig:LISTANet}.
There are $T$ layers of soft-thresholding operations in CLISTA. The convolutional dictionaries $\bm{W}_{a}$ and $\bm{W}_{s}$ are shared across layers. For the $t$-th layer, there is a soft-thresholding operator with learnable soft-thresholds: 
\begin{equation}
    \bm{g}_{t} = \mathcal{T}_{\bm{\lambda}_t} \left( \bm{g}_{t-1} + \bm{W}_{a} \otimes  ( \bm{d}^{k}_M - \bm{W}_{s} \otimes \bm{g}_{t-1} ) \right),
\end{equation}
where $\bm{W}_{a} \in \mathbb{R}^{N \times (c-h) \times r \times r}$ is the analysis convolutional dictionary, $\bm{W}_{s} \in \mathbb{R}^{(c-h) \times N \times r \times r}$ is the synthesis convolutional dictionary, and $\bm{\lambda}_t \in \mathbb{R}^{N}$ is the soft-threshold vector. 

For the first layer, $\bm{g}_{0}$ is initialized as $\bm{W}_{a} \otimes \bm{d}^{k}_M$.
At the last layer, the synthesis convolution layer $\bm{W}_s$ is used to reconstruct the denoised detail channels from the estimated sparse feature $\bm{g}_T$, \textit{i.e.}, $\hat{\bm{d}}^{k}_{M}= \bm{W}_s \otimes \bm{g}_T$.


\subsection{Noise Adaptive Network}
\label{sec:noiseAda}




To achieve robust image denoising, we propose to adapt the soft-thresholds in PUNets and CLISTA to the noise level $\sigma$ in the image. For images with larger noise levels, the soft-thresholds will be larger leading to more effective denoising.
This is similar to the model-based image denoising methods in \cite{donoho1994ideal, donoho1995adapting, chang2000adaptive, blu2007sure, elad2006image, dong2011sparsity, buades2005non, mahmoudi2005fast, dabov2007image, gu2014weighted, MSWNNW} where the soft-thresholds scale with the noise levels. For noisy image super-resolution problem, DeepAM \cite{Huang2020DeepAM} also shows that better generalization can be achieved by adjusting soft-thresholds with respect to noise level.

Different from BF-CNN which does not have bias terms and has restricted expressive power, our PUNets and CLISTA denoising network have the dynamically adjusted soft-thresholds with respect to the noise level. This gives us good generalization ability, flexibility as well as improved expressive power. 

\textbf{Spectral Norm Loss:}
When learning from noisy-clean image pairs of a single noise level $\sigma_N$, the parameters of PUNets are learned to optimize the performance of this training noise level. When the test noise level $\sigma_T < \sigma_N$, the rescaled soft-thresholds could make the PUNets change from a contractive function to an expansive function. 
We therefore use a spectral norm loss to regularize the property of the PUNets:
\begin{equation}
    \mathcal{L}_{{s}} = \frac{1}{K \cdot M \cdot J} \sum_{k=1}^{K} \sum_{m=1}^{M} \sum_{j=1}^{J} \Vert P_{m,j}^k \Vert_S + \Vert U_{m,j}^k\Vert_S,
    \label{eq:linearloss}
\end{equation}
where $\Vert\cdot\Vert_S$ denotes the spectral norm of a matrix, and  $P_{m,j}^k$ and $U_{m,j}^k$ are the effective linear filter of the $j$-th residual block on the $m$-th lifting step at the $k$-th scale when soft-thresholds are set to zeros.


The linear loss $\mathcal{L}_{{s}}$ is used to regularize the Lipschitz constant of the PUNets when $\sigma_T \rightarrow 0$, \textit{i.e.}, assuming that soft-thresholding operators in PUNets are identity functions. 
It can improve the generalization ability of WINNets to unseen small noise levels and can also stabilize the training of deep/multi-scale models by constraining the expansiveness of the PUNets.

\textbf{Orthogonal Loss:}
For CLISTA denoising network, the soft-thresholds will also be adjusted with respect to noise level. We would like to enforce $\hat{\bm{d}}^{k}_{M} = \bm{d}_M^k$ when $\sigma_T = 0$. That is, the CLISTA denoising network should be close to an identity function when there is no noise. This can be achieve when $\bm{W}_s$ and $ \bm{W}_a$ are orthogonal.
Therefore, an orthogonal loss is imposed between the synthesis and the analysis convolutional dictionary:
\begin{equation}
    \mathcal{L}_{{o}} = \Vert \bm{W}_s \otimes \bm{W}_a - \bm{\delta}  \Vert_F^2,
\end{equation}
where $\bm{\delta} \in \mathbb{R}^{(c-h) \times (c-h) \times (2r - 1) \times (2r - 1)}$ is the Kronecker delta function which takes value 1 at coordinate $(i,i,r,r)$ $\forall i \in [1, \cdots, c-h]$ otherwise 0.

\subsection{Noise Estimation Network}
\label{sec:noiseEst}

\begin{figure}
    \centering
    \includegraphics[width=0.4\textwidth]{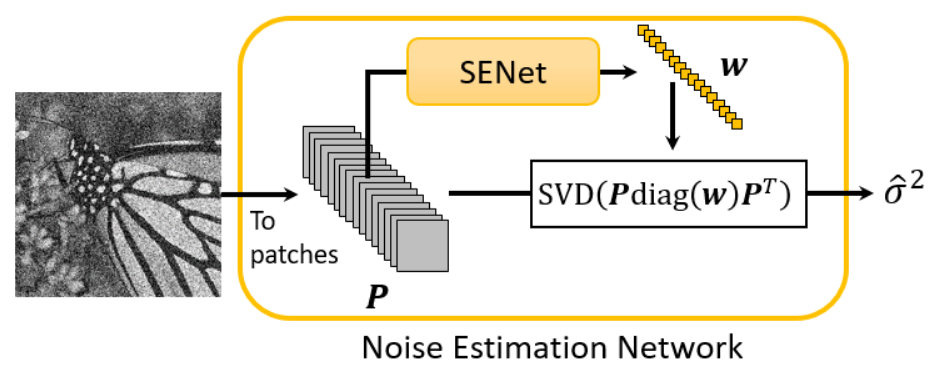}
    \caption{The proposed noise estimation network. The input noisy image is first divided into patches. For each patch, a scalar weight will be estimated using the SENet. The noise level is estimated as the smallest singular value of the weighted covariance matrix.}
    \label{fig:noiseEst}
\end{figure}

Since the PUNets are conditioned on the image noise level, we propose a simple model-inspired noise estimation network in order to achieve blind image denoising. 
The proposed noise estimation network is inspired by \cite{liu2013single} here, authors use an iterative algorithm to select low-rank patches from the input noisy image and estimate the noise level as the smallest eigenvalue of the covariance matrix of the selected low-rank patches. Compared to CNNs used to infer noise levels, \textit{e.g.}, \cite{Guo2019Cbdnet, Helou2020BUIFD}, this model-based method can provide more interpretable results and robust generalization ability.


We propose a learnable noise estimation network (NENet) which is based on the principles of \cite{liu2013single} and uses backpropagation through singular value decomposition (SVD) in PyTorch. The network is shown in Fig. \ref{fig:noiseEst}. 
The input image is divided into $\bm{P} \in \mathbb{R}^{s^2 \times N_p}$ patches where each column represents a vectorized $s \times s$ patch and $N_p$ is the number of patches. 
Instead of iteratively selecting low-rank patches, we propose to use a selection network (SENet) $\mathcal{S}: \mathbb{R}^{s^2} \rightarrow \mathbb{R}^{1}$ to estimate a scalar weight $w_i \in [0, 1]$ for each patch. 
The SENet performs a soft selection and will learn to assign large weights to low-rank patches and vice versa.
Given the estimated weights and the patches, the noise variance is estimated as the minimum singular value of the weighted covariance matrix of the patch matrix:
\begin{equation}
    \hat{\sigma}^2 = \frac{1}{\sum_{i=1}^{N_p} w_i} {\sigma}_{\text{min}},
\end{equation}
where ${\sigma}_{\text{min}}$ denotes the smallest singular value of the weighted covariance matrix $\bm{P} \text{diag}(\bm{w}) \bm{P}^T$ with $\bm{w}=[w_1, \cdots, w_{N_p}]^T$.

The SENet is composed of blocks with 1-D convolution layer and ReLU non-linearity without bias term, followed by a global averaging pooling and a Sigmoid function to map the weight in the range of $[0,1]$. The parameters of SENet can be learned using backpropagation.
For the noise estimation network, the MSE noise loss is used:
\begin{equation}
    \mathcal{L}_{{n}} = \frac{1}{2N} \sum_{i=1}^N \left( \sigma_i - \hat{\sigma}_i \right)^2,
\end{equation}
where $\sigma_i$ and $\hat{\sigma}_i$ denotes the ground-truth noise level and the estimated noise level, respectively.


\subsection{Training Details}

WINNet can learn from noisy-clean image pairs of a single noise level or a range of noise levels as in \cite{zhang2017beyond, zhang2018ffdnet, Guo2019Cbdnet}. The average mean squared error (MSE) between the restored image and the clean image is used as the reconstruction loss:
\begin{equation}
    \mathcal{L}_{{r}} = \frac{1}{2N} \sum_{i=1}^{N} \Vert \bm{x}_i - \bm{\hat{x}}_i \Vert_2^2,
\end{equation}
where $\bm{x}_i$ and $\bm{\hat{x}}_i$ denotes the input clean image and the denoised image, respectively.

The overall training objective of our WINNet is:
\begin{equation}
    \mathcal{L}_{all} = \mathcal{L}_{{r}} + \lambda_{1} \mathcal{L}_{{s}} + \lambda_{2} \mathcal{L}_{{o}},
\end{equation}
where $\lambda_{1}$, and $\lambda_{2}$ are regularization parameters.

\section{Experimental Results} \label{sec:results}

In this section, we perform experiments to show the properties and validate the effectiveness of the proposed WINNet. We will first introduce the experimental settings, visualize the learned transform and the sparsity-driven denoising network results, and finally perform comparisons with other methods.

\subsection{Implementation Details}

\begin{figure*}[t]
    \centering
    \centering
	\hspace*{\fill}
	\subfigure[Level 1, coarse channel.]{
		\includegraphics[width=0.35\textwidth]{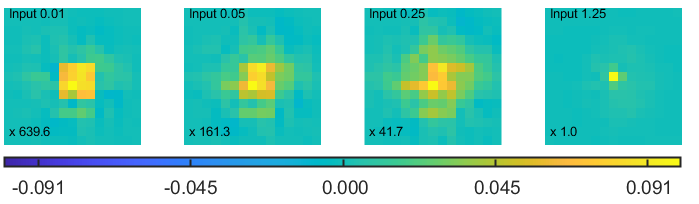}}
	\hfill
	\subfigure[Level 1, detail channel 5.]{
		\includegraphics[width=0.35\textwidth]{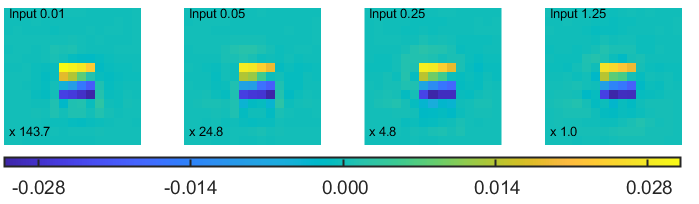}}
	\hspace*{\fill}
	
	\hspace*{\fill}
	\subfigure[Level 1, detail channel 10.]{
		\includegraphics[width=0.35\textwidth]{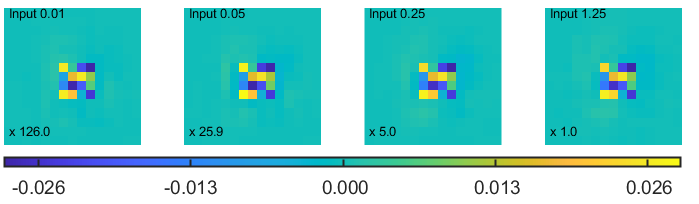}}
	\hfill
	\subfigure[Level 1, detail channel 14.]{
		\includegraphics[width=0.35\textwidth]{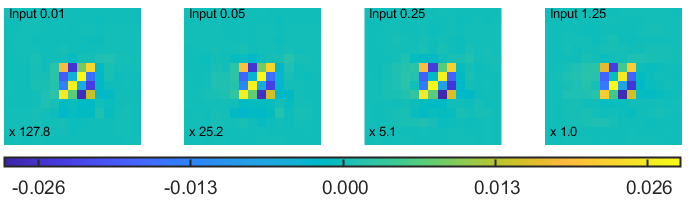}}
	\hspace*{\fill}
	
	\hspace*{\fill}
	\subfigure[Level 2, coarse channel.]{
		\includegraphics[width=0.35\textwidth]{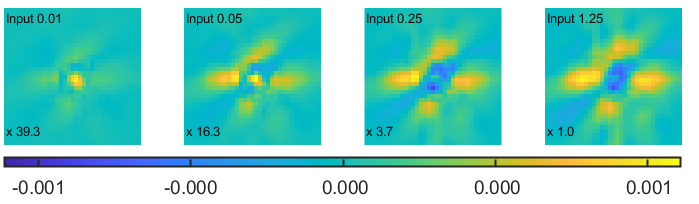}}
	\hfill
	\subfigure[Level 2, detail channel 5.]{
		\includegraphics[width=0.35\textwidth]{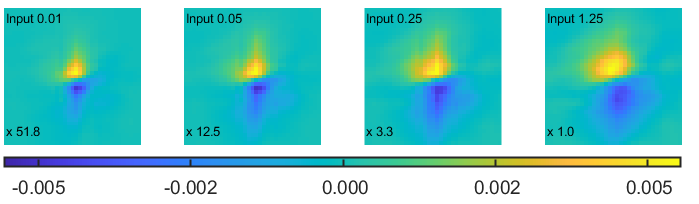}}
	\hspace*{\fill}
	
	\hspace*{\fill}
	\subfigure[Level 2, detail channel 10.]{
		\includegraphics[width=0.35\textwidth]{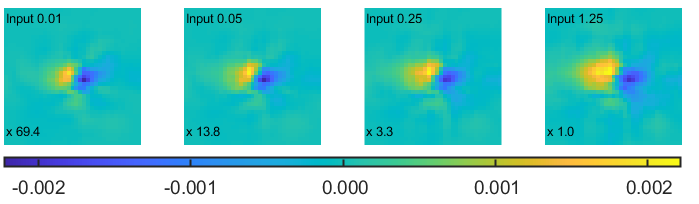}}
	\hfill
	\subfigure[Level 2, detail channel 14.]{
		\includegraphics[width=0.35\textwidth]{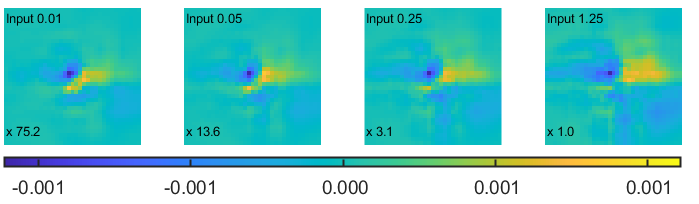}}
	\hspace*{\fill}
	
    \caption{Visualization results of the basis functions learned by LINNs. The amplitude of the non-zero value $v_{nz}$ is shown on the upper left corner of each figure. For better visualization, the maximum absolute value in each visualized basis function has been rescaled by a factor which is shown on the lower left corner of each figure.
    }
    \label{fig:visualizeBasis}
\end{figure*}

We follow the training and evaluation settings in \cite{zhang2017beyond} to perform experiments. Before training and testing, the clean images are normalized to $[0,255]$. The noisy images are obtained by adding additive white Gaussian noise to the clean image with respect to Eq. (\ref{eq:noise}) with variance $\sigma^2$. 

\subsubsection{WINNet Configuration}
For default setting, the convolution kernel $\bm{K}_s \in \mathbb{R}^{c \times 1 \times p \times p}$ of the splitting operator are reshaped from a DCT transform matrix $\bm{T} \in \mathbb{R}^{p^2 \times p^2}$ (\textit{i.e.} the $i$-th row of $\bm{T}$ is the reshaped $i$-th $p \times p$ filter in $\bm{K}_s$). We set $c=p^2$ and $p=4$. The convolution kernel for the merging operator can be constructed in a similar manner. Since by default $p=4$, there will be 1 coarse channel and 15 detail channels.

The number of update and predict network pairs is set to $M = 4$, and the number of residual blocks in each PUNet is set to $J=4$. The number of feature channels in PUNet is set to 32 and the spatial filter size in SepConv layers is set to $q = 5$.
For CLISTA denoising network, there are $T=3$ layers with 64 channels. The support of the spatial filter for $\bm{W}_a$ and $\bm{W}_s$ is set to $r = 3$.

\subsubsection{Training and Testing settings}
The 400 training images from the Berkeley segmentation dataset (BSD) \cite{roth2009fields} of size $180 \times 180$  are used for training. The training patch size is $40 \times 40$, and the number of patches for training is $2.3 \times 10^5$.

For training with a single noise level, three noise levels are considered, \textit{i.e.}, $\sigma_N = 15, 25$ and $50$. 
For blind image denoising scenario, the training noise level $\sigma_N$ is uniformly drawn from $[0,55]$.
The regularization parameter for spectral norm loss and orthogonal loss is set to $\lambda_1 = 0.1$ and $\lambda_2 = 10$, respectively. 
The spectral norm loss is evaluated once every 10 iterations.
As discussed in Section \ref{sec:noiseAda}, when training blind image denoising models, all the soft-thresholds will be rescaled with respect to the training noise level.


The weights of the convolution layers are initialized using the Kaiming initialization method \cite{he2015delving}.
The stochastic gradient descent with Adam optimizer \cite{Kingma2014} is used for training with initial learning rate $lr=1 \times 10^{-3}$ and $\beta=(0.9, 0.999)$.
The total number of training epochs is set to 50 and the learning rate decays from $1 \times 10^{-3}$ to $1 \times 10^{-4}$ at the 30-th epoch. The batch size $N$ is set to 32. 

The testing images include the 12 images from \textit{Set12} \cite{zhang2017beyond}, and the 68 natural images from the \textit{BSD68} \cite{roth2009fields}. PSNR is used as the evaluation metric.

\vspace{-0.3cm}

\subsection{Visualizing the Intermediate Results}

In this section, we will visualize the output produced by components of a WINNet with $K=2$ levels which is trained on data with noise level $\sigma_N = 25$. 

\subsubsection{Elementary Building Blocks Learned by WINNet}

\begin{figure*}[t]
    \centering
    
	\hspace*{\fill}
	\subfigure[$\bm{c}_M^1$.]{
		\includegraphics[height=0.13\textwidth]{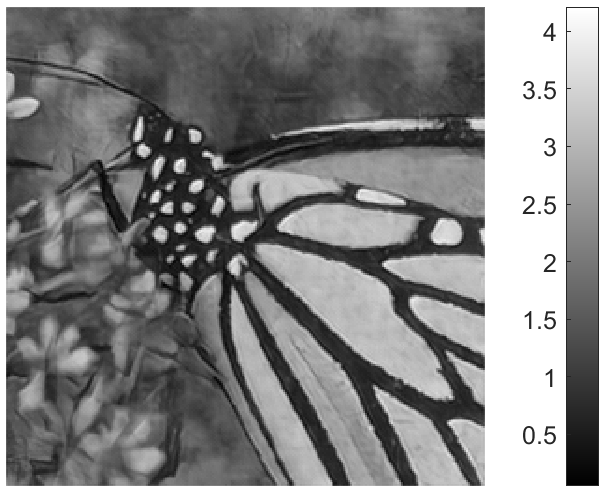}}
	\hfill
	\subfigure[$\bm{d}_M^1(1)$.]{
		\includegraphics[height=0.13\textwidth]{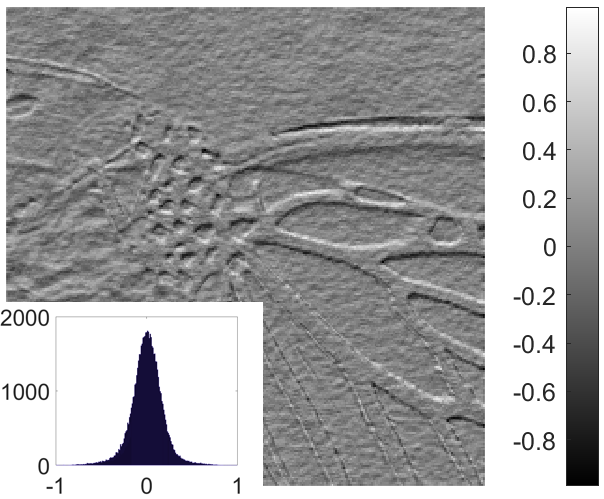}}
	\hfill
	\subfigure[$\hat{\bm{d}}_M^1(1)$.]{
		\includegraphics[height=0.13\textwidth]{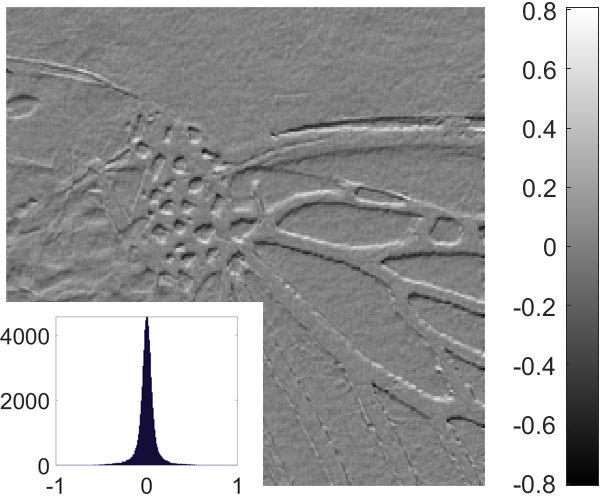}}
	\hfill
	\subfigure[$\bm{d}_M^1(5)$.]{
		\includegraphics[height=0.13\textwidth]{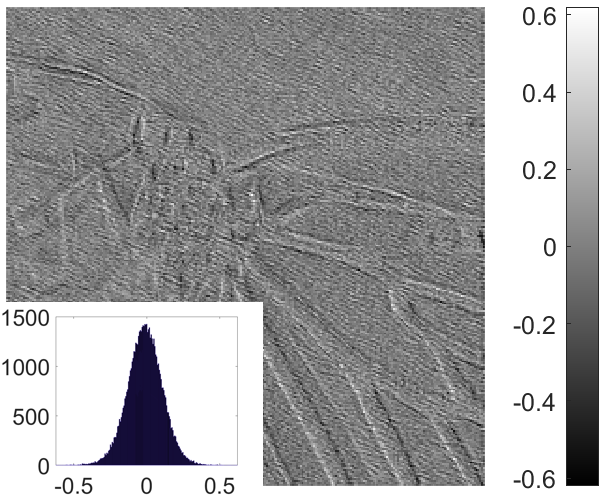}}
	\hfill
	\subfigure[$\hat{\bm{d}}_M^1(5)$.]{
		\includegraphics[height=0.13\textwidth]{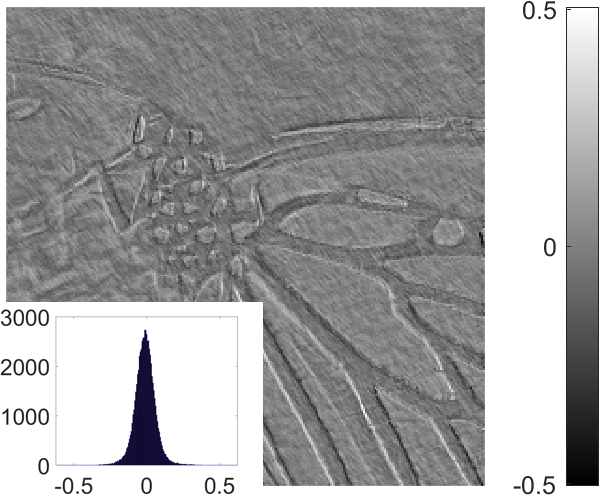}}
	\hspace*{\fill}

	\hspace*{\fill}
	\subfigure[$\bm{c}_M^2$.]{
		\includegraphics[height=0.13\textwidth]{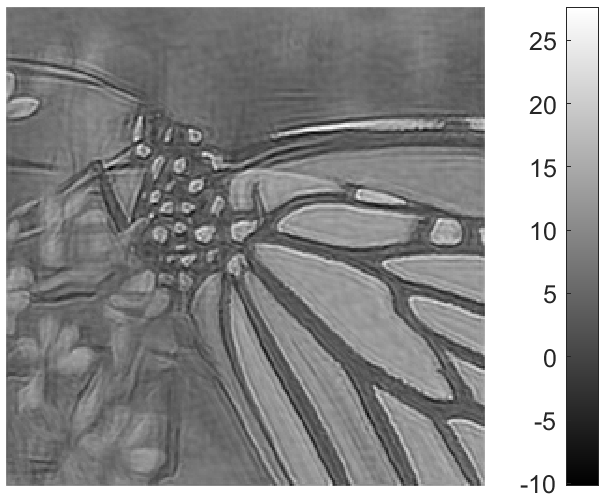}}
	\hfill
	\subfigure[$\bm{d}_M^2(1)$.]{
		\includegraphics[height=0.13\textwidth]{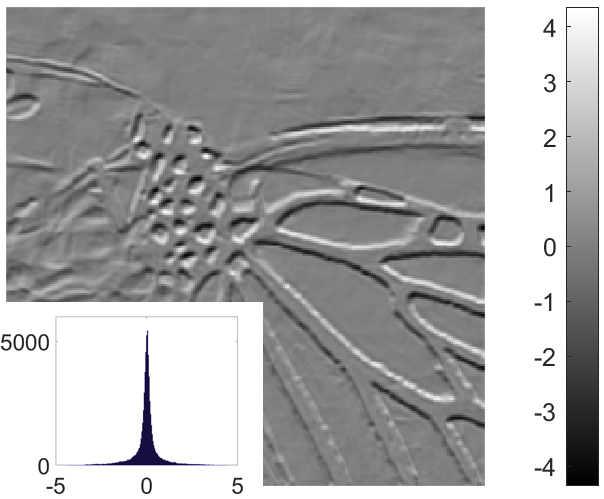}}
	\hfill
	\subfigure[$\bm{d}_M^2(1) - \hat{\bm{d}}_M^2(1)$.]{
		\includegraphics[height=0.13\textwidth]{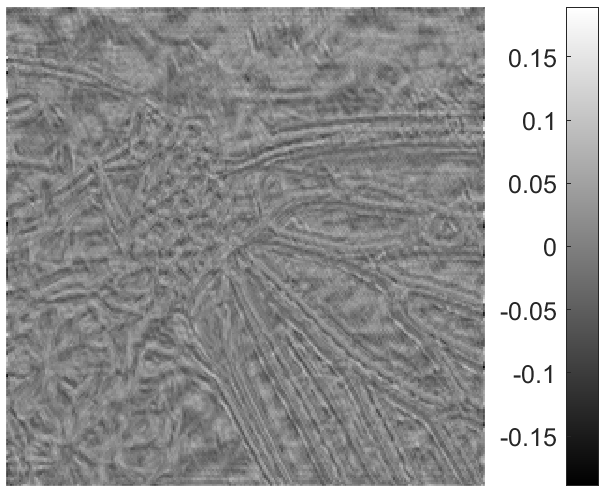}}
	\hfill
	\subfigure[$\bm{d}_M^2(5)$.]{
		\includegraphics[height=0.13\textwidth]{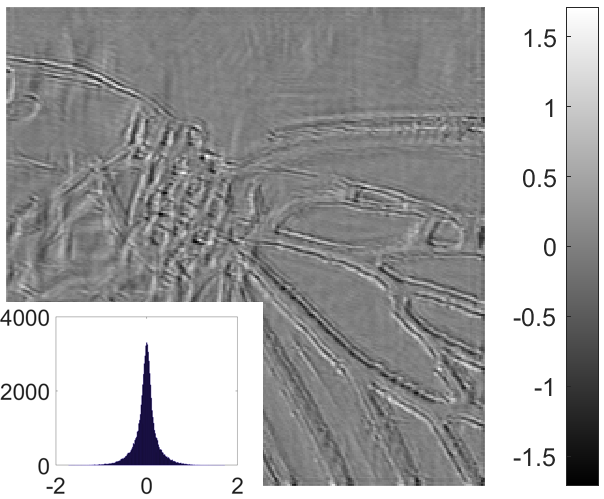}}
	\hfill
	\subfigure[$\bm{d}_M^2(5) - \hat{\bm{d}}_M^2(5)$.]{
		\includegraphics[height=0.13\textwidth]{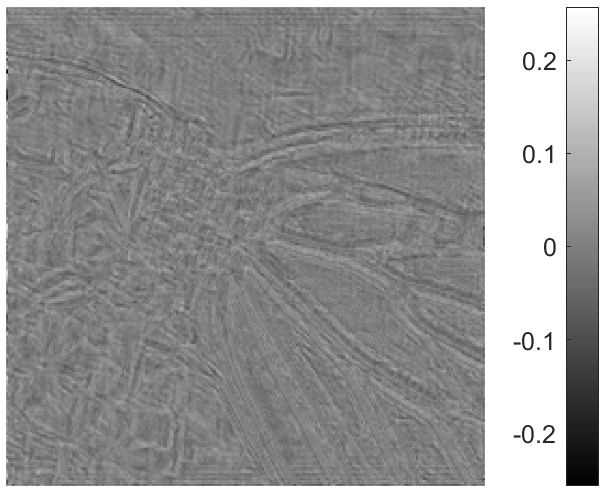}}
	\hspace*{\fill}


    \caption{Visualization of the coarse and detail channels. (a) and (f) show the coarse component at two different scales. (b) - (e) and (g) - (j) show the detail channels before and after denoising by CLISTA denoising network in WINNet. Each sub-figure shows the feature map of a detail channel and the corresponding histogram.
    ($\bm{d}_M^k(j)$ denotes the $j$-th channel of the detail part.)
    }
    \label{fig:visualize_lvl1}
\end{figure*}

The proposed invertible network is inspired by the lifting scheme \cite{sweldens1998lifting, daubechies1998factoring} in wavelets, therefore, it would be meaningful to visualize the learned elementary atoms. 

The basis functions or elementary atoms of the wavelet transform can be visualized by setting to zero all the wavelet coefficients with the exception of one coefficient and then by reconstructing the corresponding signal with the synthesis filter bank. In analogy with the wavelet case, we set all the feature maps in $( {\bm{c}}^{k}_M, \hat{\bm{d}}^{k}_M )$ to zero with the exception of the center pixel on one of the feature maps, and then reconstruct the image using the backward pass of LINN.
As the PUNets represent highly non-linear functions and may have different responses for signals with varying amplitudes, the elementary atoms is visualized using different non-zero values.
Since our transform is highly non-linear, it is incorrect to call the reconstructed functions ``basis functions''or elementary atoms, however, to keep the intuition that what we are visualizing is a variation of the basis functions related to linear transforms, with a slight abuse, we keep calling them ``elementary atoms''.

Fig. \ref{fig:visualizeBasis} shows the atoms corresponding to different channels at level 1 and level 2 in WINNet. For better visualization, the maximum absolute value in each visualized basis function has been rescaled by a factor shown on the lower left corner of each figure. 
Fig. \ref{fig:visualizeBasis} (a) - (d) show the elementary atoms of 4 representative channels from level 1 LINN. We can see that the basis functions have compact support. For the coarse channel, the basis function gradually changes to a delta function when the amplitude increases from 0.01 to 1.25. For the detail channels, the shape of the elementary atoms only slightly changes when we increase the amplitude of the input pixel. 
Fig. \ref{fig:visualizeBasis} (e) - (h) show the atoms of 4 representative channels from level 2 LINN. Different from what we observed at level 1, the basis functions in level 2 have larger support, and the shapes change from concentrated to spread.

Both level 1 and level 2 functions have non-linear responses to the input amplitude, while the atoms of the level 2 have much larger support compared to that of the level 1.
The different support size of the functions at level 1 and level 2 is possibly due to their different functionality in WINNet, and is also a consequence of the dilation of the filter support discussed before and depicted in Fig. \ref{fig:atrous}.

\begin{figure}
    \centering
    \hspace*{\fill}
    \subfigure[Clean image.]{
		\includegraphics[width=0.1\textwidth]{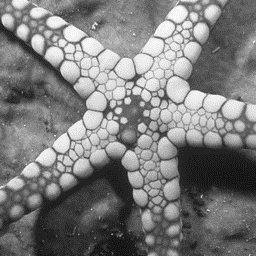}}
	\hfill
	\subfigure[$\sigma=1$.]{
		\includegraphics[width=0.1\textwidth]{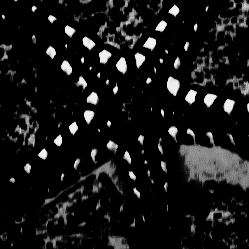}}
	\hfill
	\subfigure[$\sigma=20$.]{
		\includegraphics[width=0.1\textwidth]{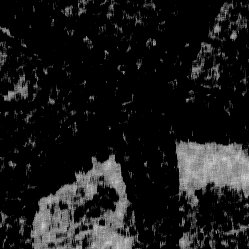}}
	\hfill
	\subfigure[$\sigma=60$.]{
		\includegraphics[width=0.1\textwidth]{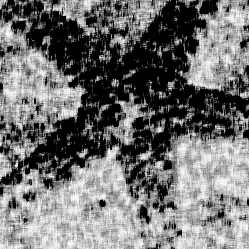}}
	\hspace*{\fill}
	
	\hspace*{\fill}
	\subfigure[The noise level estimation generated by NENet which is trained using data with $\sigma_N \in \text{[0,55]}$ (marked in green).]{
		\includegraphics[width=0.42\textwidth]{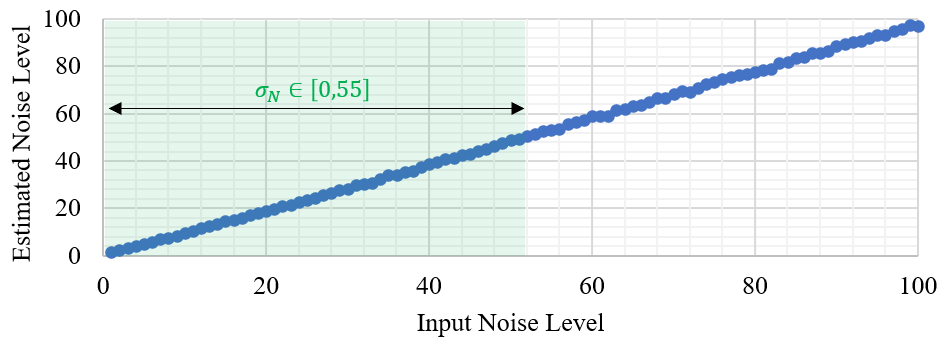}}
	\hspace*{\fill}
	
    \caption{Visualization of the selected patches for noise level estimation and the generalization ability of the NENet.}
    \label{fig:NENet}
\end{figure}

\begin{table*}[]
    \caption{The model size and average PSNR (dB) of different \textit{NON-BLIND} image denoising methods tested on \textit{BSD68} and \textit{Set12} dataset with noise level $\sigma=15,25,50$. (The best and the second best result in each column is in bold and with underline, respectively.)}
    \label{tab:nonblind}
    \centering
    \begin{tabular}{|l|l| C{1.8cm}||C{1.8cm}|C{1.8cm}|C{1.8cm}|}
\hline 
Dataset                & Methods                    & Model Size          & $\sigma=15$ & $\sigma=25$ & $\sigma=50$ \\ \hline \hline 
\multirow{9}{*}{\textit{BSD68}} 
& BM3D \cite{dabov2007image}    & -                   & 31.07         & 28.57         & 25.63         \\  
& WNNM \cite{gu2014weighted}    & -                   & 31.37         & 28.83         & 25.87         \\  
& EPLL \cite{zoran2011learning} & -                   & 31.21         & 28.68         & 25.67         \\  
& TNRD \cite{chen2016trainable} & $27 \times 10^3$  & 31.42         & 28.92         & 25.97         \\  
& DnCNN-S \cite{zhang2017beyond} & $556 \times 10^3$ & \textbf{31.71}         & 29.20         & {26.22}         \\  
& BF-CNN \cite{mohan2020BFCNN} & $556 \times 10^3$ & 31.58    & 29.08     & 26.12       \\ 
& FFDNet \cite{zhang2018ffdnet}  & $556 \times 10^3$ & 32.63       & 29.19         & \underline{26.29}      \\  
& WINNet (1-scale)               & $173 \times 10^3$ & {31.69}         & \underline{29.22}      & \underline{26.29}         \\  
& WINNet (2-scale)               & $347 \times 10^3$ & \underline{31.70}         & \textbf{29.24}         & \textbf{26.31}              \\ \hline \hline 

\multirow{9}{*}{\textit{Set12}} 
& BM3D \cite{dabov2007image}    & -                   & 32.37         & 29.97         & 26.72         \\  
& WNNM \cite{gu2014weighted}    & -                   & 32.70         & 30.26         & 27.05         \\  
& EPLL \cite{zoran2011learning} & -                   & 32.14         & 29.69         & 26.47         \\  
& TNRD \cite{chen2016trainable} & $27 \times 10^3$  & 32.50         & 30.06         & 26.81         \\  
& DnCNN-S \cite{zhang2017beyond}  & $556 \times 10^3$ & \textbf{32.83}         & {30.40}       & 27.16         \\  
& BF-CNN \cite{mohan2020BFCNN}  & $556 \times 10^3$ & 32.67       & 30.25         & 27.03      \\  
& FFDNet \cite{zhang2018ffdnet}  & $556 \times 10^3$ & 32.75       & \underline{30.43}         & \textbf{27.32}      \\  
& WINNet (1-scale)           & $173 \times 10^3$ & \underline{32.82}         & \underline{30.43}      & {27.25}         \\  
& WINNet (2-scale)           & $347 \times 10^3$ & \textbf{32.83}         & \textbf{30.47}         & \underline{27.29}              \\ \hline \hline

    \end{tabular}
\end{table*}

\subsubsection{Noisy and Denoised Feature Maps} 

The CLISTA denoising network is the only non-invertible component in WINNet, and is responsible for removing the feature components corresponding to noise. By visualizing the feature maps before and after the denoising network, we can further gain insights into the workings of the proposed WINNet.

In Fig. \ref{fig:visualize_lvl1}, we show the feature maps of an exemplar \textit{Butterfly} image from \textit{Set12}. Each sub-figure shows the feature map of a channel and the corresponding histogram.
Fig. \ref{fig:visualize_lvl1} (a) (b) (d) show the feature map of 3 exemplar channels output from the level 1 LINN in WINNet. We can see that the coarse channel feature map looks like a natural image but is still with artifacts and the detail channel feature maps before denoising contain both noise and some image contents and the histogram is spread.
Fig. \ref{fig:visualize_lvl1} (c) - (e) show the corresponding detail channels after denoising. We can see that the noise has been significantly reduced, the edges are sharper and the histograms become more concentrated around the origin. 
This indicates that noise has been effectively removed by CLISTA denoising network.

In Fig. \ref{fig:visualize_lvl1} (f), the coarse channel feature map contains high-frequency enhanced features. For the detail channels, there is no significant differences on the feature maps before and after the CLISTA denoising network since the 1-st level coarse channel feature $\bm{c}_M^1$ (shown in Fig. \ref{fig:visualize_lvl1}(a)) mainly contains minor artifacts. Therefore, we only show the feature maps before denoising in Fig. \ref{fig:visualize_lvl1} (g) (i), and also include the difference of the feature maps before and after denoising in Fig. \ref{fig:visualize_lvl1} (h) (j). We can observe that the 2-nd level CLISTA makes minor modifications to the feature maps.


\subsubsection{Noise Estimation Network}

The proposed noise estimation network is based on the idea that the Gaussian noise level can be estimated as the smallest eigenvalue of the low-rank patches. Therefore, NENet would have highly interpretable results and strong generalization ability.

Fig. \ref{fig:NENet} (a) shows an exemplar clean image, and (b) - (e) shows the region of the selected low-rank patches by the selection network (SENet) for different noise levels, respectively. We can see that SENet tends to select different regions for noise estimation. For low noise level, SENet carefully selects smooth regions, while for high noise level SENet avoids highly textured regions and selects more patches. The results are consistent with that in \cite{liu2013single} which uses a heuristic and iterative algorithm to select low-rank patches. Fig. \ref{fig:NENet} (c) further shows the noise estimation results for images with noise level $\sigma_T \in [0,100]$. We can find that the proposed NENet is able to provide highly accurate noise estimation for images with $\sigma_T \in [0,100]$ though the training data is only with noise level $\sigma_N \in [0,55]$.

\subsection{Comparison with Other Methods}

We compare the proposed WINNet with several state-of-the-art image denoising algorithms including the model-based methods: BM3D \cite{dabov2007image}, WNNM \cite{gu2014weighted}, EPLL \cite{zoran2011learning}, and the learning-based methods: DnCNN \cite{zhang2017beyond}, FFDNet \cite{zhang2018ffdnet}, BUIFD \cite{Helou2020BUIFD}, BF-CNN \cite{mohan2020BFCNN}. 
The model-based methods can handle images with arbitrary noise levels. DnCNN \cite{zhang2017beyond} and BUIFD \cite{Helou2020BUIFD} can perform blind image denoising when the test noise level is within the range of the training noise levels, BF-CNN \cite{mohan2020BFCNN} can handle unseen noise levels, while the FFDNet \cite{zhang2018ffdnet} requires as input a noise map of the test noise level.

\begin{figure}
    \centering
    \includegraphics[width=0.4\textwidth]{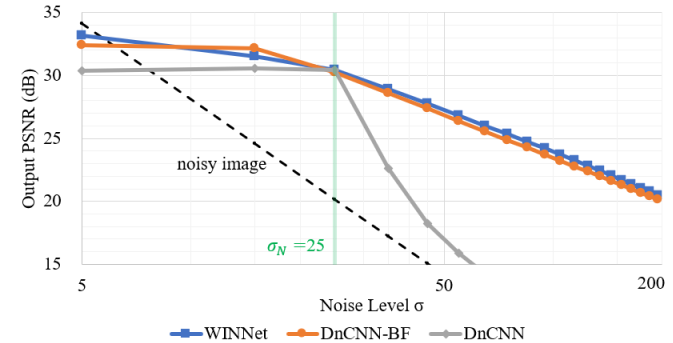}
    \caption{The noise robustness of different methods which are trained on noise level $\sigma_N=25$ and tested on \textit{Set12} with noise level $\sigma_T \in [5, 200]$.}
    \label{fig:noiseAdapt}
\end{figure}






\begin{table*}[]
    \caption{The average PSNR (dB) of different \textit{BLIND} image denoising methods trained on \textit{BSD400} dataset with noise level $\sigma \in [0,55]$ and tested on \textit{BSD68} and \textit{Set12} dataset with noise level $\sigma \in [5,145]$. (The best and the second best result in each column is in bold and with underline, respectively.)}
    \label{tab:blind}
    \centering
    \begin{tabular}{|l|l||  C{1.2cm}|C{1.2cm}|C{1.2cm}||C{1.2cm}|C{1.2cm}|C{1.2cm}|C{1.2cm}|C{1.2cm}|C{1.2cm}|C{1.2cm}|}
\hline 
Dataset & Methods  & $\sigma=5$ & $\sigma=25$ & $\sigma=45$ & $\sigma=65$ & $\sigma=85$ & $\sigma=105$ & $\sigma=125$ & $\sigma=145$ \\ 
\hline \hline 

\multirow{4}{*}{\textit{BSD68}} 
& DnCNN-B \cite{zhang2017beyond}    & \underline{37.75} & \textbf{29.15} & \underline{26.62} & 23.00 & 16.07 & 13.19 & 11.68 & 10.79        \\ 
& BUIFD \cite{Helou2020BUIFD}       & 37.41 & 28.76 & 25.61 & 23.07 & 18.81 & 15.98 & 14.45 & 13.52        \\  
& BF-CNN \cite{mohan2020BFCNN}      & 37.73 & 29.11 & 26.58 & \underline{25.12} & \underline{24.10} & \underline{23.33} & \underline{22.70} & \underline{22.18}        \\  
& WINNet (1-scale)                  & \textbf{37.82} & \underline{29.13} & \textbf{26.66} & \textbf{25.23} & \textbf{24.23} & \textbf{23.46} & \textbf{22.81} & \textbf{22.23}        \\  
\hline \hline

\multirow{4}{*}{\textit{Set12}} 
& DnCNN-B \cite{zhang2017beyond}    & \underline{37.88} & \textbf{30.38} & \underline{27.68} & 23.52 & 15.95 & 13.18 & 11.78 & 10.92        \\  
& BUIFD \cite{Helou2020BUIFD}       & 37.34 & 30.18 & 27.01 & 24.27 & 19.41 & 16.28 & 14.66 & 13.73        \\  
& BF-CNN \cite{mohan2020BFCNN}      & 37.81 & \underline{30.33} & 27.58 & \underline{25.83} & \underline{24.54} & \underline{23.55} & \underline{22.74} & \underline{22.07}        \\  
& WINNet (1-scale)                  & \textbf{38.22} & \underline{30.33} & \textbf{27.72} & \textbf{26.03} & \textbf{24.77} & \textbf{23.76} & \textbf{22.94} & \textbf{22.24}        \\  
\hline \hline
                       
    \end{tabular}
\end{table*}
\begin{figure*}
    \centering
    \hspace*{\fill}
    \subfigure[Input noisy image with different noise levels]{
		\includegraphics[width=0.8\textwidth]{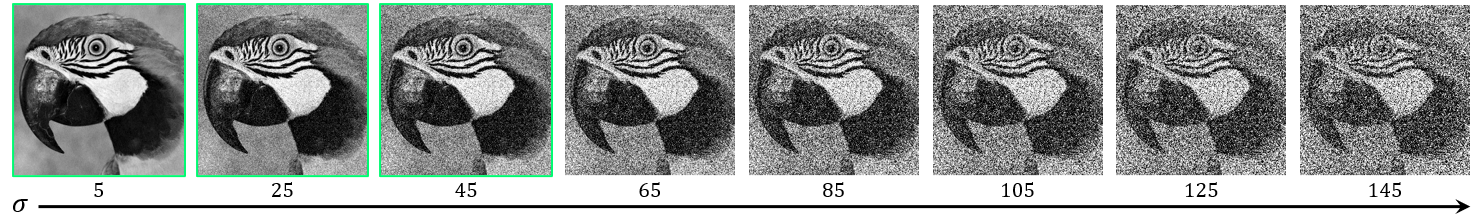}}
	\hspace*{\fill}
	
	\hspace*{\fill}
    \subfigure[The corresponding image denoising results by WINNet.]{
		\includegraphics[width=0.8\textwidth]{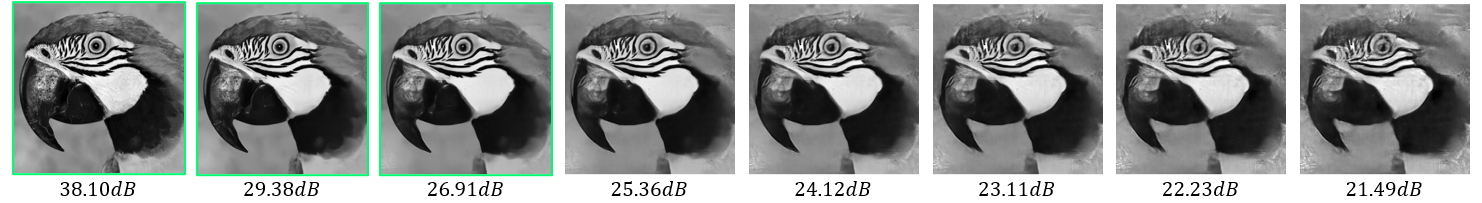}}
	\hspace*{\fill}
	
    \caption{Exemplar noisy images and blind image denoising results of blind WINNet whose training noise levels $\sigma \in [0,55]$. The first row shows the noisy image with different noise levels, and the second row shows the denoising results. The images within the training noise levels are outlined with green.}
    \label{fig:non-blind}
\end{figure*}

\subsubsection{Non-blind Image Denoising}

Table \ref{tab:nonblind} shows the comparison results of different non-blind image denoising methods evaluated on three noise levels (\textit{i.e.,} $\sigma=15,25,50$). All the learning-based methods learn from training samples with the correct noise level.
From the table, we can see that the proposed WINNets achieve better performance than the model-based methods \textit{i.e.}, BM3D \cite{dabov2007image}, WNNM \cite{gu2014weighted}, and EPLL \cite{zoran2011learning} by a large margin, and our proposed WINNets also outperforms the deep learning based methods including TNRD \cite{chen2016trainable}, DnCNN-S \cite{zhang2017beyond}, BF-CNN \cite{mohan2020BFCNN} and FFDNet \cite{zhang2018ffdnet}. 
We also note that the model size of WINNet (1-scale) is only around 30\% of the model size of DnCNN, and WINNet (1-scale) achieves comparable or better results than DnCNN-S. With one more level decomposition, WINNet (2-scale) further improves the WINNet (1-scale).

Without the bias terms, BF-CNN \cite{mohan2020BFCNN} has stronger generalization ability to testing images with unseen noise levels than DnCNN-S (we will show this in Fig. \ref{fig:noiseAdapt}), while in Table \ref{tab:nonblind} the performance of BF-CNN is around 0.15 dB lower than that of the DnCNN-S. This suggests that the generalization ability is improved at the cost of reduced expressiveness.

WINNet has strong generalization ability to images with unseen noise levels as well.
Although WINNet only sees training image pairs with a fixed noise level $\sigma_N$, its parameters can be adjusted to adapt to unseen noise levels. When the testing noise level $\sigma_T \geq \sigma_N$, all the soft-thresholds in PUNets and in CLISTA denoising networks are rescaled by a factor $\sigma_T / \sigma_N$. When $\sigma_T < \sigma_N$, only the soft-thresholds in CLISTA denoising networks are rescaled by a factor $\sigma_T / \sigma_N$.

Fig. \ref{fig:noiseAdapt} shows the performance of the proposed WINNet, BF-CNN and DnCNN-S which are trained on noise level $\sigma_N=25$, while are tested on testing noise levels $\sigma_T \in [5,195]$. 
We can see that the performance of DnCNN remains similar when $\sigma_T \leq \sigma_N$ and quickly deteriorates otherwise. For BF-CNN and WINNet, they can well generalize to testing images with noise levels $\sigma_T \neq \sigma_N$. 
When $\sigma_T \geq \sigma_N$, WINNet achieves improved gain compared to BF-CNN. When $\sigma_T =15$, the performance of WINNet is inferior to BF-CNN, but when $\sigma_T =5$, WINNet achieves around 0.8 dB higher PSNR than BF-CNN.

\subsubsection{Blind Image Denoising} 

For blind image denoising, the comparison methods include DnCNN-B \cite{zhang2017beyond}, BUIFD \cite{Helou2020BUIFD}, and BF-CNN \cite{mohan2020BFCNN}. The depth of DnCNN-B and BF-CNN are increased form 17 to 20 and their number of parameters is around $660 \times 10^3$. For WINNet, we use the 1-scale model for comparison which has around $173 \times 10^3$ parameters. The number of parameters for NENet is only around $6 \times 10^3$.
The training data for all methods is \textit{BSD400} with AWGN $\sigma_N \in [0, 55]$.

Table \ref{tab:blind} shows the testing results of different methods evaluated with images with noise level $\sigma \in [5,145]$. We can see that the performance of DnCNN-B \cite{zhang2017beyond} is highly competitive when the testing noise level is within the range of the training noise level, while quickly deteriorates otherwise. 
BUIFD \cite{Helou2020BUIFD} consists of a noise level CNN, a prior CNN and a fusion network which are all based on DnCNN architecture. Its total number of parameters is around $1186 \times 10^3$. With explicit noise level learning, BUIFD shows stronger generalization ability towards unseen noise levels compared to DnCNN-B. 
By removing all bias terms in DnCNN, BF-CNN \cite{mohan2020BFCNN} is able to well generalize beyond the training noise levels, however, it is slightly less effective when $\sigma \in [0,55]$ compared to DnCNN.
With the exception of $\sigma=25$, WINNet consistently outperforms all the other methods.

Fig. \ref{fig:non-blind} further shows the noisy image with different noise levels and the blind image denoising results by the proposed WINNet. We can see that the proposed WINNet is able to achieve robust denoising not only within the training noise level range (marked in green) but also beyond the training noise levels.

\begin{figure*}
    \centering
    \includegraphics[width=0.8\textwidth]{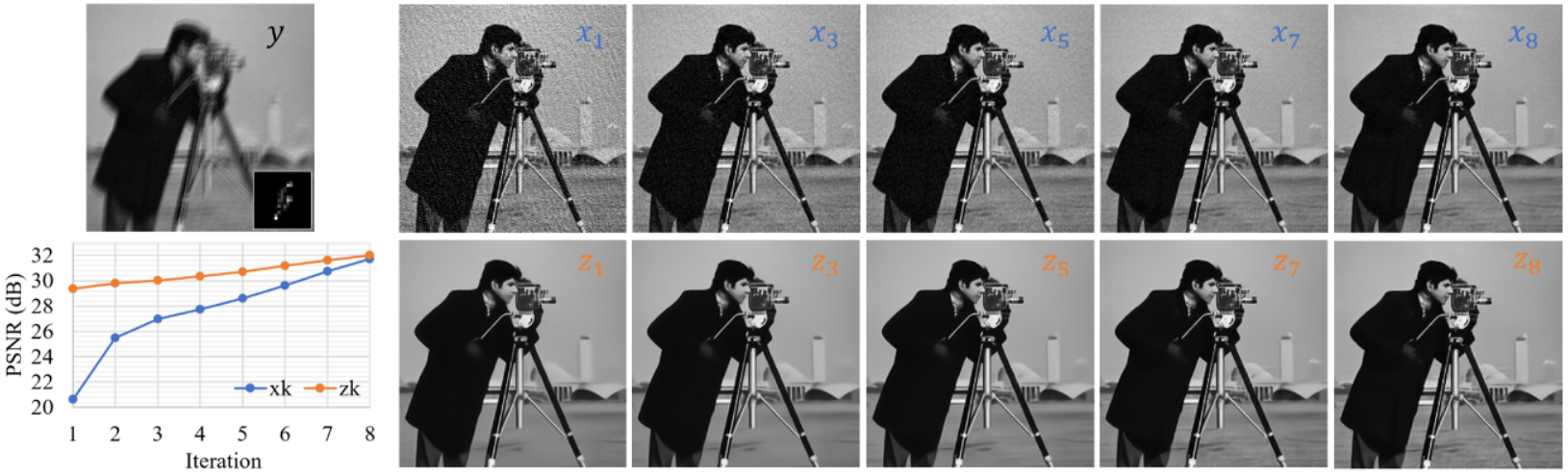}
    \caption{The input blurred image $\bm{y}$ with the blurring kernel $\bm{k}$, the convergence curve and the visualization results of $\bm{x}_k$ and $\bm{z}_k$ at different iterations. The blurring kernel is the first kernel of size $19 \times 19$ from \cite{levin2009understanding} and the noise level is 2.55 (1\%).}
    \label{fig:deblur_CMan}
\end{figure*}

\begin{table*}[]
\centering
\caption{The average PSNR (dB) of different image deblurring methods evaluated on \textit{Set12} with kernels from \cite{levin2009understanding} and noise level $2.55$ (1\%). }

\label{tab:deblur}

\begin{tabular}{|c||C{1.2cm}|C{1.2cm}|C{1.2cm}|C{1.2cm}|C{1.2cm}|C{1.2cm}|C{1.2cm}|C{1.2cm}|}

\hline 
\multirow{3}{*}{\textit{kernel}}                               & 1 & 2 & 3 & 4   & 5 & 6 & 7 & 8   \\ 
&  \includegraphics[width=0.03\textwidth]{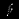} 
& \includegraphics[width=0.03\textwidth]{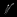} 
& \includegraphics[width=0.03\textwidth]{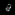} 
& \includegraphics[width=0.03\textwidth]{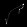}   
& \includegraphics[width=0.03\textwidth]{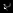} 
& \includegraphics[width=0.03\textwidth]{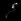} 
& \includegraphics[width=0.03\textwidth]{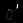} 
& \includegraphics[width=0.03\textwidth]{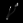}   \\ \hline \hline
EPLL \cite{zoran2011learning}       & 31.47 & 31.00 & 31.23 & 29.80   & \textbf{32.47} & 32.28 & 31.12 & 30.53   \\ 
IRCNN \cite{zhang2017learning}      & 32.26 & 31.65 & 30.87 & \textbf{31.76}   & 31.82 & 31.99 & 31.10 & 31.07   \\
Proposed                            & \textbf{32.37} & \textbf{32.04} & \textbf{32.03} & 31.52   & \textbf{32.47} & \textbf{33.17} & \textbf{32.02} & \textbf{31.87}   \\
\hline 
\end{tabular}
\end{table*}

\vspace{-0.2cm}

\subsection{Application on Image Deblurring}

With the plug-and-play technique, image denoisers can be applied to solve general image restoration problems \cite{burger2012image, chen2016trainable, zhang2017beyond, zhang2018ffdnet, Guo2019Cbdnet}, for example, image deblurring. In this case, the goal is to recover a sharp image $\bm{x}$ from the blurred and noisy observation $\bm{y} = \bm{k} \otimes \bm{x} + \bm{n}$ where $\bm{k}$ is the blurring kernel and $\bm{n} \sim  \mathcal{N}(0,\sigma^2)$ represents the measurement noise with variance $\sigma^2$.
The image deblurring task can be formulated as the following optimization problem:
\begin{equation}
    \bm{x} = \arg \underset{\bm{x}}{\min} \frac{1}{2 \sigma^2} \Vert \bm{y} - \bm{k} \otimes \bm{x} \Vert_2^2 + \lambda \Phi(\bm{x}),
    \label{eq:deblur}
\end{equation}
where $\Phi(\cdot)$ is a prior term and $\lambda$ is the regularization parameter.
With half-quadratic splitting \cite{zhang2017learning}, the image deblurring problem can be solved by iteratively optimizing two sub-problems:
\begin{align}
    \bm{x}_k =& \arg \underset{\bm{x}}{\min} \Vert \bm{y} - \bm{k} \otimes \bm{x} \Vert_2^2 + \frac{\lambda \sigma^2}{\beta^2} \Vert \bm{x} - \bm{z}_{k-1} \Vert_2^2,\\
    \bm{z}_k =& \arg \underset{\bm{z}}{\min} \frac{1}{2 \beta^2} \Vert \bm{z} - \bm{x}_{k} \Vert_2^2 + \Phi(\bm{z}),
\end{align}
where $\beta=\sqrt{\lambda / \mu}$ is a hyper-parameter and can be interpreted as the noise level if the $\bm{z}$ sub-problem is treated as Gaussian denoising on $\bm{x}_{k}$.

In \cite{zhang2017learning}, a CNN-based Gaussian denoiser is used to solve the $\bm{z}$ sub-problem. The hyper-parameter $\lambda$ is set to be fixed during iterations, while $\mu$ is set to exponentially decay from a large value to the given noise level $\sigma$ with a fixed iteration number. Since the proposed NENet is an effective noise level estimator and WINNet can denoise images with noise beyond training noise levels, we propose to use WINNet for image deblurring. 
The $\bm{x}$ sub-problem can be solved with closed-form solution with the estimated noise level $\beta_k$ by NENet. The $\bm{z}$ sub-problem can be solved using the proposed blind WINNet with noise level $2 \beta_k$ (perform denoising with a stronger strength to ensure convergence). 
With the proposed robust NENet and WINNet, we can achieve image deblurring without accessing the noise levels and using the pre-defined regularization parameters; $\lambda$ is the only free parameter and is set to 0.23 as in \cite{zhang2017learning}.
Algorithm \ref{alg:PnPWINNet} illustrates the plug-and-play image deblurring algorithm with the proposed WINNet.


\begin{algorithm}[t]
    \SetAlgoLined
    \textbf{Input:} Input image $\bm{y}$, kernel $\bm{k}$, parameter $\lambda$;
    
    \textbf{Initialize:} $\bm{z}_0 = \bm{y}$, $\beta_0 = \text{NENet}(\bm{z}_0)$, $\beta_1 = 10 \times \beta_0$, $k=1$;\\

    \While{$\beta_k > \beta_0$}{
        $\bm{x}_k = \arg \underset{\bm{x}}{\min} \Vert \bm{y} - \bm{k} \otimes \bm{x} \Vert_2^2 + \frac{\lambda \beta_0^2}{\beta_{k}^2} \Vert \bm{x} - \bm{z}_{k-1} \Vert_2^2$;\\
        
        $\beta_{k+1} = \text{NENet}(\bm{x}_k)$;\\
        $\bm{z}_k = \text{WINNet}(\bm{x}_k, 2 \beta_{k+1})$;\\
        
        $k = k + 1$;\\
    }
    \textbf{Output:} Deblurred image $\bm{x} = \bm{z}_{k-1}$.
 \caption{Plug-and-Play image deblurring with blind WINNet.}
 \label{alg:PnPWINNet}
\end{algorithm}

Table \ref{tab:deblur} shows the average PSNR (dB) of the EPLL \cite{zoran2011learning}, IRCNN \cite{zhang2017learning} and the proposed method evaluated on \textit{Set12} with 8 different kernels from \cite{levin2009understanding} and noise level 2.55. We can see that the proposed method is able to achieve highly competitive performance.
Fig. \ref{fig:deblur_CMan} shows an exemplar image deblurring process of the proposed method on image \textit{Cameraman} which is blurred using the first kernel from \cite{levin2009understanding}. We can see that the proposed method converges after 8 iterations and the PSNR of $\bm{x}_k$ and $\bm{z}_k$ consistently improves and finally reaches a similar result.

\section{Conclusions} \label{sec:conclude}

In this paper, we have proposed a wavelet-inspired invertible network (WINNet). It consists of $K$ levels of lifting inspired invertible neural network (LINN) and sparsity-driven denoising networks. LINNs are designed to mimic the nice properties of wavelet transform and are used as a non-linear redundant transform with perfect reconstruction property. For image denoising task, the sparsity-driven denoising network is used to remove the noise in the detail parts of the transform coefficients and the denoising network can be adjusted to adapt to unseen noise levels. Together with a model-inspired noise estimation network, the proposed blind WINNet can achieve robust blind image denoising results beyond the training noise levels. The flexibility of WINNet has also been demonstrated on the image deblurring task.




\bibliographystyle{IEEEtran}
\bibliography{bibs}

\end{document}